\documentclass[english]{article}
\pdfoutput=1
\usepackage[latin9]{inputenc}
\usepackage{geometry}
\usepackage{babel,amsmath,amssymb,authblk,graphicx}
\usepackage[numbers,sort&compress]{natbib}
\usepackage[unicode=true,
 bookmarks=true,bookmarksnumbered=false,bookmarksopen=false,
 breaklinks=false,pdfborder={0 0 1},backref=false,colorlinks=false]
 {hyperref}
\usepackage{multirow}
\usepackage{tabularx}
\usepackage{breakurl,arydshln,bm,url,pifont}
\usepackage[dvipsnames]{xcolor}

\usepackage[normalem]{ulem}
\usepackage{cancel}
\usepackage{diagbox}

\mathchardef\Theta="7102
\def\sss#1{{\scriptscriptstyle #1}}

\def\ssr#1{{\sss{\rm #1}}}
\def\ie{{\it i.e.\/}}

\def\frac#1#2{{\textstyle{#1 \over #2}}}
\def\half{\frac{1}2}
\def\yd{^\dagger}
\def\nd{^{\vphantom{\dagger}}}
\def\ns{^{\vphantom{*}}}
\def\np{^{\vphantom{\prime}}}

\def\vth{\vartheta}

\def\intl{\int\limits}
\def\Bk{{\bm k}}
\def\BF{{\bm F}}
\def\BA{{\bm A}}
\def\BB{{\bm B}}

\def\Bn{{\bm n}}

\def\Bl{{\bm l}}
\def\Ba{{\bm a}}
\def\Bb{{\bm b}}
\def\BR{{\bm R}}
\def\Ht{{\hat t}}
\def\Hf{{\hat f}}

\def\Btheta{{\bm\theta}}
\def\Bvth{{\bm\vartheta}}

\def\Bpi{{\bm \pi}}
\def\Bzero{\boldsymbol 0}
\def\eps{\epsilon}

\def\CO{{\cal O}}
\def\MZ{\mathbb{Z}}
\def\TT{{\textsf{T}^2}}
\def\TTT{{\textsf{T}^3}}
\def\iTT{\int\limits_\TT} 
\def\iTTT{\int\limits_\TTT}
\def\Lap{\Lambda^\parallel}
\def\chib{{\bar\chi}}
\def\nbar{{\bar n}}
\def\dtil{{\tilde\delta}}

\def\blangle{{\big\langle}}
\def\brangle{{\big\rangle}}
\def\Rep{\textsf{Re}}
\def\Imp{\textsf{Im}}
\def\pz{\partial}
\def\Tra{\textsf{Tr}\,}
\def\det{\textsf{det}\,}

\def\sket#1{{|  \, #1 \,  \rangle}}

\def\sexpect#1#2#3{{\langle \, #1 \, | \,  #2  \, | \, #3 \, \rangle}}

\def\Braket#1#2{{\Big\langle \, #1\, \Big| \,#2 \,\Big\rangle}}

\makeatletter

\usepackage{caption}
\usepackage{subcaption}

\makeatother

\begin{document}

\title{Fluctuation of Chern Numbers in a Parametric Random Matrix Model}
\author[1]{Hung-Hwa Lin}
\author[1]{Wei-Ting Kuo}
\author[1]{Daniel P. Arovas}
\author[1]{Yi-Zhuang You}
\affil[1]{Department of Physics, University of California at San Diego, La Jolla, California 92093, USA}
\date{}

\maketitle
\begin{abstract}
    Band-touching Weyl points in Weyl semimetals give rise to many novel characteristics, 
    one of which the presence of surface Fermi-arc states that is topologically protected. 
    The number of such states can be computed by the Chern numbers at different momentum slices, which fluctuates with changing momentum and depends on the distribution of Weyl points in the Brillouin zone. 
    For realistic systems, it may be difficult to locate the momenta at which these Weyl points and Fermi-arc states appear.
    Therefore, we extend the analysis of a parametric random matrix model proposed by Walker and Wilkinson to find the statistics of their distributions. 
    Our numerical data shows that Weyl points with opposite polarities are short range correlated, and the Chern number fluctuation only grows linearly for a limited momentum difference before it saturates. 
    We also find that the saturation value scales with the total number of bands. 
    We then compute the short-range correlation length from perturbation theory, and derive the dependence of the Chern number fluctuation on the momentum difference, 
    showing that the saturation results from the short-range correlation. 
\end{abstract}

\tableofcontents{}

\section{Introduction}

The band-touching points of Weyl semimetals, called Weyl points, are responsible for its novel characteristics, such as the surface Fermi arc states \cite{wan2011topological}, chiral anomaly \cite{aji2012adler,hosur2013recent,son2013chiral,xiong2015signature}, and the anomalous Hall effect \cite{yang2011quantum}. 
The existence of these Weyl points are topologically protected in three-dimensional materials with either time-reversal or inversion symmetry broken, 
and cannot be gapped out by perturbing the system \cite{murakami2007phase}. 
They serve as monopoles of Berry curvature inside the bulk Brillouin zone, each with a polarity defined by the sign of the Berry flux through an enclosing surface. 
Also, they result in non-zero Chern numbers for some of the Brillouin zone slices, which manifest as surface Fermi-arc states present only in between Weyl points \cite{wan2011topological}. 
These have been most commonly observed in three-dimensional solids \cite{armitage2018weyl}, while also appearing in other systems such as the spectra of polyatomic molecules \cite{faure2000topological}, nanomagnets \cite{wernsdorfer2005quantum}, quantum transport systems \cite{leone2008cooper}, and more recently multichannel Josephson junctions \cite{riwar2016multi}. 
In another context, band-touching points can appear in the analysis of a two-dimensional system with a tunable parameter, where the topological invariant can only change with the parameter tuned to such point \cite{walker1995universal}. 

Since the locations of these Weyl points can be difficult to solve for realistic systems, one may ask if any universal statistics of can be obtained for the location distribution. 
The statistics of the Chern number defined on a 2D surface, which fluctuates when the surface is displaced, is also of interest, 
as it gives the number of Fermi-arc states measurable in experiments.
Therefore, motivated by the idea that complex systems can be well-described by random matrix ensembles \cite{mehta2004random}, 
we extend the analysis of a parameterized random matrix model,
proposed by Walker and Wilkinson to describe the statistics of the Weyl points and Chern number fluctuations \cite{walker1995universal,wilkinson1993GOE,austin1992statistical}.
In their work, 
the average density of Weyl points in the parameter space has been computed, exhibiting a scaling relation with respect to the total number of bands, 
and the fluctuation of the Chern numbers in the limit of small displacement of the surface analyzed, which showed linear dispersion.
Later works have shown the correlations of the degeneracy points to exhibit perfect screening \cite{wilkinson2004screening} and analyzed the correlations of the Berry curvature \cite{gat2021correlations}. 
A modified version of this model comprised of random matrices obeying the Bogoliubov-de-Gunnes mirror symmetry was also considered to describe the Weyl points in multichannel josephson junctions \cite{barakov2021abundance}. 

Building on these results, we make further investigation that yields numerical data on the correlations of Weyl point locations and the Chern number fluctuation for finite displacement of the surface, and propose an analytical model based on perturbation theory that reproduces these results.
The opposite-polarity Weyl points show short range correlation, while the Chern number fluctuation saturates at a plateau after initial linear dispersion for small displacement, 
and we explain how the former leads to the latter.
In addition, our data shows a scaling relation for the height of the saturation plateau with respect to the total number of bands.
We then compute the short-range correlation length from perturbation theory, which also scales with the total number of bands. 
Finally, assuming that the correlation function is exponential, we analytically compute the Chern number fluctuation and the scaling of the plateau, which matches the numerical data. 

This work is organized as follows: in Sec.\ref{sec:model}, we review the parameterized random matrix model proposed in \cite{walker1995universal} and illustrate the connection between the Weyl point correlations and the Chern number fluctuation;
numerical results are shown in Sec.\ref{sec:numerical}, and in Sec.\ref{sec:analytical} we present analytical calculations for the correlations and the Chern number fluctuation; we conclude with a summary and discussion Sec.\ref{sec:discussion}. 

\section{Model and Background}\label{sec:model}

In this section, we review backgrounds on the Weyl points and Fermi arc states for generic multi-band electronic systems, 
and discuss a specific parametric random matrix Hamiltonian we adopt to models them. 
Generically, such a system can in  be described by a single-particle Hamiltonian $H \left( \theta_1, \theta_2, \theta_3 \right)$,
where $\Btheta=(\theta_1,\theta_2,\theta_3)$ parameterizes the electron crystal momentum in a condensed matter setting, as
we shall describe in detail below.  Weyl points appear as degeneracies at isolated $\Btheta$ values, 
and Fermi arc states may be detected by computing Chern numbers on constant $\theta_3$ slices of the momentum space. 
We will illustrate how the distribution of Weyl points and the fluctuation of Chern numbers with $\theta_3$ are closely related, both of which we wish to understood. 
However, since they can be difficult to compute for generic systems, we describe a parametric random matrix Hamiltonian adopted from previous work \cite{walker1995universal}, 
which will allow numerical and analytical calculation of the statistics of these quantities. 

The underlying model is taken to be a general tight-binding model of the form
\begin{equation}
H=\sum_{\Bn,\Bn'}\sum_{a,a'} t\ns_{aa'}(\Bn-\Bn')\,c\yd_{a\np}(\Bn)\,c\nd_{a'}(\Bn')\quad,
\end{equation}
where $t\ns_{aa'}(\Bn-\Bn')$ is the hopping amplitude for an electron in orbital $a'$ in unit cell $\BR'=\sum_{j=1}^d n'_j\,\Ba\np_j$
to hop to the orbital $a$ in unit cell $\BR=\sum_{j=1}^d n\np_j\,\Ba\np_j$\,, with $d$ the dimension of space and 
$\{\Ba\ns_1,\ldots,\Ba\ns_d\}$ a set of elementary direct lattice vectors; we will focus on the case $d=3$.  In Fourier space, we have
\begin{equation}
H=\sum_\Btheta\sum_{a,a'} \Ht\ns_{aa'}(\Btheta)\,c\yd_{a\np}(\Btheta)\,c\nd_{a'}(\Btheta)\quad,
\end{equation}
where $\Btheta$ is the dimensionless wavevector, $N\ns_{\rm c}\to\infty$ the number of unit cells,
$c\nd_a(\Btheta)=N_{\rm c}^{-1/2}\sum_\Bn c\ns_a(\Bn)\,e^{i\Bn\cdot\Btheta}$, and
$\Ht\ns_{ab}(\Btheta)=\sum_\Bn t\ns_{ab}(\Bn)\,e^{i\Bn\cdot\Btheta}$.  The physical wavevector is given by 
$\Bk=\sum_{j=1}^d\theta\ns_j\Bb\ns_j/2\pi$, where $\{\Bb\ns_1,\ldots,\Bb\ns_d\}$ are the elementary reciprocal lattice vectors,
with $\Ba\ns_i\cdot\Bb\ns_j=2\pi\delta\ns_{ij}$\,.  Hermiticity requires $\Ht(\Btheta)=\Ht\yd(\Btheta)$ for each $\Btheta$, which
entails $t(-\Bn)=t\yd(\Bn)$.  We require that the matrices $t(\Bn)$ are Gaussianly random distributed, with
\begin{equation}
\blangle t\ns_{ab}(\Bn)\,t\ns_{cd}(\Bn')\brangle = f(\Bn)\,\delta\ns_{\Bn+\Bn',\Bzero}\,\delta\ns_{ad}\,\delta\ns_{bc}\quad,
\end{equation}
where each $f(\Bn)=f(-\Bn)$ is real and non-negative.  The distribution function of the $t(\Bn)$ matrices is then given by
\begin{equation}
P\big[\{t(\Bn)\}\big]= C\,{\prod_\Bn}' \exp\Bigg\{\!-{\Tra\big[t\yd(\Bn)\,t(\Bn)\big]\over 2 f(\Bn)}\Bigg\}\quad,
\end{equation}
where the prime indicates that only one of $\{\Bn,-\Bn\}$ is included in the product for all $\Bn$.  This entails
\begin{equation}
\blangle \Ht\ns_{ab}(\Btheta)\,\Ht\ns_{cd}(\Btheta')\brangle = \Hf(\Btheta-\Btheta')\,\delta\ns_{ad}\,\delta\ns_{bc}\quad,
\end{equation}
where $\Hf(\Bvth)=\sum_\Bn f(\Bn)\,e^{-i\Bn\cdot\Bvth}=\Hf^*(\Bvth)=\Hf(-\Bvth)$ is the structure factor of our generalized
Walker-Wilkinson ensemble.  We normalize our distribution by demanding $\Hf(\Bzero)=1$, hence for each $\Btheta$ the $\Ht(\Btheta)$
matrices are distributed according to the GUE with 
$\blangle \Ht\ns_{ab}(\Btheta)\,\Ht\ns_{cd}(\Btheta)\brangle =\delta\ns_{ad}\,\delta\ns_{bc}$\,. We will denote $\hat{t}(\Btheta)$ as $H(\Btheta)$ in what follows.

\subsection{Weyl Points and Chern Numbers}\label{WPCN}

To discuss Weyl points and Chern numbers, we first reiterate that
a three-dimensional multi-band electronic systems can be described by a single-particle Hamiltonian $H( \theta_1, \theta_2, \theta_3)$. 
Associated with each point $\Btheta$ is an energy spectrum $E_n(\Btheta)$ and eigenstates $\sket{\psi_n(\Btheta)}$. 
Now consider the possibility of making neighboring band $n$ and $n+1$ degenerate by tuning the momentum $\Btheta$. 
Since it is well known that three parameters are required to achieve this \cite{von1929uber}, the occurrence of such momenta $\Btheta$ will appear as isolated points since the momentum space is precisely three-dimensional. 
For the same reason, these points cannot be gapped out by perturbations. 
This is how the band-touching Weyl-points appear in a generic system. 

For the Chern numbers, which detect the number of Fermi-arc states \cite{wan2011topological}, 
we need the Berry connection $A_{n,\mu}(\Btheta)=\sexpect{\psi_n(\Btheta)}{\pz_\mu}{\psi_n(\Btheta)}$
for each energy band $n$, where $\pz_\mu=\pz/\pz \theta_\mu$\,.
A Chern number can thus be computed by integrating the Berry curvature over a closed surface, 
where we consider constant $\theta_3$ surfaces, 
\begin{equation}
C\nd_n(\theta\ns_3)={1\over 2\pi}\intl_0^{2\pi} \!d\theta\ns_1\intl_0^{2\pi} \!d\theta\ns_2\>
\eps^{\mu\nu}\,\Braket{{\pz\psi_n(\Btheta)\over\pz \theta_\mu}}
{{\psi_n(\Btheta)\over\pz \theta_\nu}}
\label{eq:C_n_def}
\end{equation}
It is associated with the $n^{\text{th}}$ energy band, and $\mu,\nu=1,2$ denote the $\theta_{1}$ and $\theta_2$ direction in the three-dimensional parameter space. It is sometimes more physical to consider the sum of Chern numbers up to a certain $n^{\text{th}}$ energy level , 
\begin{equation}
S\nd_n(\theta\ns_3)\equiv\sum_{j=1}^n C_{j}(\theta\ns_3)\quad,
\label{Seqn}
\end{equation}
which could be the total Chern number for the occupied states below the Fermi energy. 

\begin{figure}
\centering
\begin{subfigure}{\textwidth}
    \centering
    \hspace{0.4cm}\includegraphics[scale=0.7]{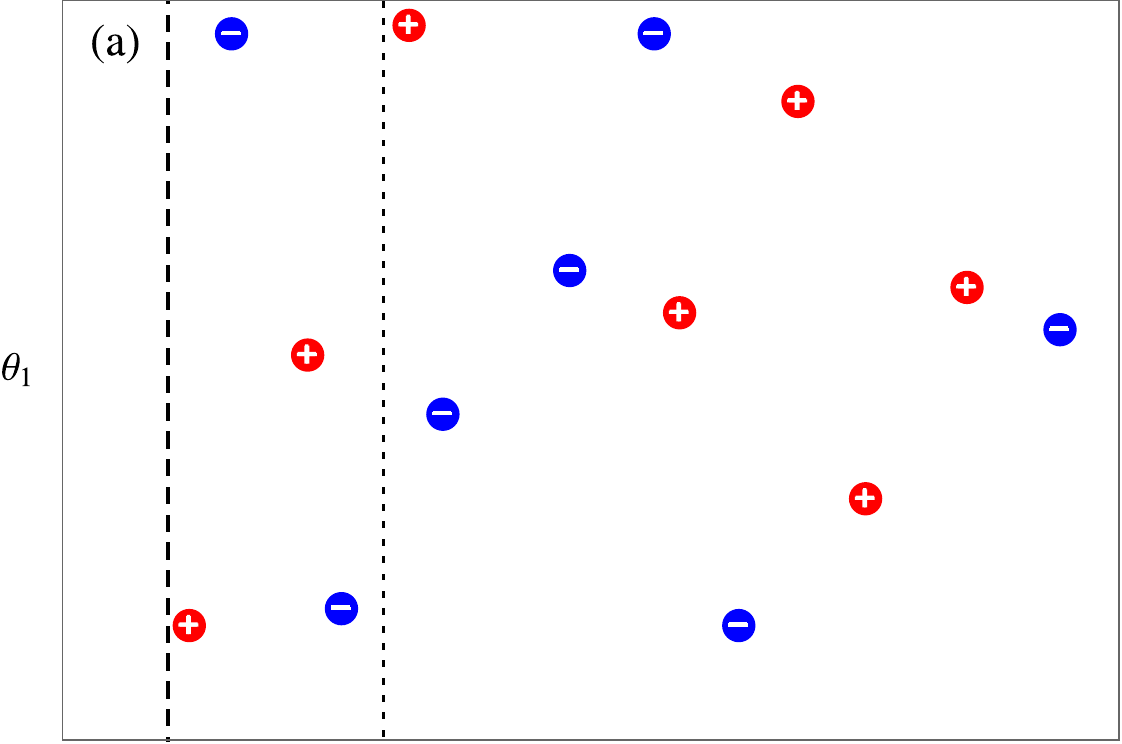}
    \phantomcaption{\label{fig:deg_2d_randomly_2d}}
\end{subfigure}
\\
\begin{subfigure}{\textwidth}
    \centering
    \includegraphics[scale=0.7]{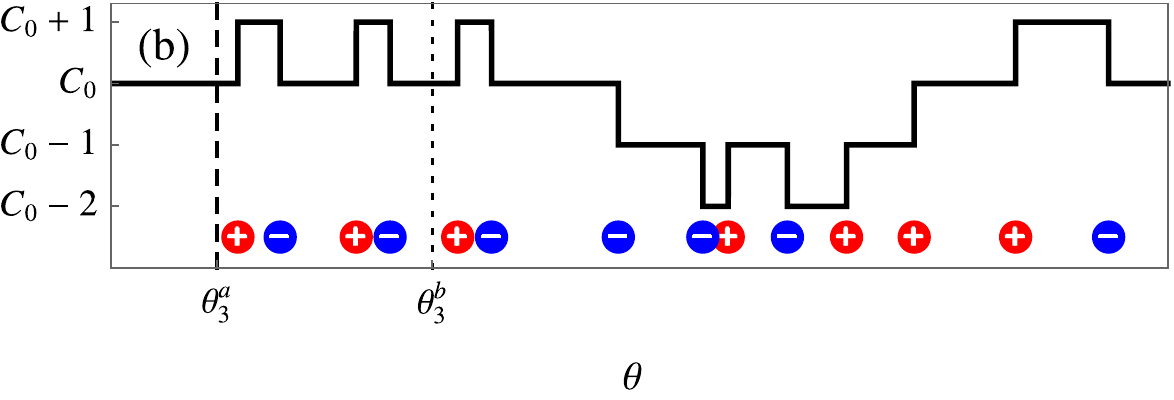}
    \phantomcaption{\label{fig:deg_2d_randomly_1d}}
\end{subfigure}
\caption{Weyl points and the change of Chern numbers. (a) the positive (red) and negative (blue) Weyl points in the $\Btheta$ space and the integration planes used to compute $C(\theta\ns_3)$ at $\theta_3=\theta_3^{a,b}$; the $\theta_2$ direction is omitted for ease of plotting. 
(b) the Weyl point distribution projected to the $\theta_3$ axis, where the Weyl points coincide with the change in the Chern number; $C_0$ denotes the initial value. \label{fig:deg_2d_randomly-1} }
\end{figure}

Since the Chern numbers only change at band-touching points because of topological protection, 
their dependences on $\theta_3$ are closely related to the distribution of Weyl points. 
To illustrate, Fig. \ref{fig:deg_2d_randomly-1}
shows Weyl points\footnote{As the degeneracy of a general Hermitian matrix is co-dimension 3,
the degeneracies in the momentum space should be point like in general. } 
distributed in the momentum space, and the constant $\theta_3$ planes (2-tori)
over which the Berry curvature is integrated to compute the Chern number. 
For example, the integration plane for $C\ns_n(\theta\np_3)$ and $S\ns_n(\theta\np_3)$
is represented by the dotted line in Fig. \ref{fig:deg_2d_randomly-1}.
Moving the plane past Weyl point would increase (decrease) the Chern number by $1$, which determines its polarity to be positive (negative). 
The $\theta_3$ coordinates of the Weyl points coincide with the those $\theta_3$ at which the Chern number changes, as shown in Fig.\ref{fig:deg_2d_randomly_1d}, 
which bears apparent similarity to random walk. 
Denoting the signed density of monopoles (Weyl points) between band $n$ and $n+1$ as $\rho_n(\Btheta)=\rho_n^+(\Btheta)-\rho_n^-(\Btheta)$,
the Chern number change is exactly the total signed number of monopoles between the plane $\theta_3=0$ and $\theta_3=\chi$,\footnote{A given monopole, which is to say an energy degeneracy between levels $n$ and $n+1$, results in a
net transfer of Chern index between these bands.  Thus a positively `charged' monopole Weyl would increase the Chern number of band $n$ and
decrease that of band $n+1$.} we have
\begin{equation}
\Delta C_n(\chi) \equiv C_n(\chi)-C_n(0) \\
= \intl_0^\chi\!d\theta_3\>\big[\lambda_n(\theta\ns_3)-
\lambda_{n-1}(\theta\ns_3)\big] \quad,
\label{DeltaCn}
\end{equation}
where $\lambda_n(\theta\ns_3)=\sum_\sigma \sigma\,\lambda^\sigma_n(\theta\ns_3)$, with
\begin{equation}
\lambda_n^\sigma(\theta\ns_3)=\intl_0^{2\pi} \!d\theta_1 \intl_0^{2\pi} \!d\theta_2
\ \rho_n^\sigma(\theta_1,\theta_2,\theta_3)
\label{lde}
\end{equation}
the one-dimensional monopole (Weyl point) density.  
Similarly, the Fermi-level Chern number is given by
$\Delta S_n(\chi)=S_n(\chi)-S_n(0)$.


\subsection{Parametric GUE model}

As we are interested in generic features of Weyl points and Chern numbers, we eschew attempts to model
real materials and instead settle on a convenient model, which is that proposed by Walker and Wlikinson \cite{walker1995universal},
\begin{equation}
H(\Btheta)= {1\over\sqrt{3}} \sum_{r=1}^3\Big[\cos \theta\ns_r\, H^{(2r-1)}+
\sin \theta\ns_r\, H^{(2r)}\Big]\quad,
\label{eq:GUE_paramed}
\end{equation}
where each of $\{H^{(1)},\ldots,H^{(6)}\}$ are independently drawn from the
Gaussian unitary ensemble (GUE) of $M\times M$ matrices, with variances
\begin{equation}
\blangle(\Rep\, H_{ab})^2\brangle=1+\delta_{ab} \qquad,\qquad
\blangle(\Imp\, H_{ab})^2\brangle=  1-\delta_{ab}\quad.
\end{equation}
The statistics of $H(\Btheta)$ and the derivative $\pz H/\pz \theta_\alpha$, for each $\Btheta$, are GUE as well \cite{walker1995universal}. Therefore, computation at and perturbation around each momentum value $\Btheta$ follow from well-known results of the GUE. The complex matrix $H(\Btheta)$ depends on three parameters, hence according to the Wigner-von Neumann theorem we expect point degeneracies in the $3$-torus $(\theta_1,\theta_2,\theta_3)$, with each $\theta_\alpha\in[0,2\pi)$.  We call the parameterized
ensemble of random matrices $H(\Btheta)$ the Walker-Wilkinson ensemble (WWE).

The structure factor $\Hf(\Bvth)$ for the WWE is given by
\begin{equation}
\Hf\ns_\ssr{WWE}(\Bvth)=\frac{1}{3}\big(\cos\vth\ns_1 + \cos\vth\ns_2 + \cos\vth\ns_3\big)\quad;
\end{equation}
note the normalization $\Hf(\Bzero)=1$.  We shall also discuss the properties of a one-parameter extension of the WWE, given by
\begin{equation}
H(\Btheta)=\sqrt{1-\alpha}\,H^{(0)} + {\sqrt{\alpha\over 3}} \sum_{r=1}^3\Big[\cos \theta\ns_r\, H^{(2r-1)}+
\sin \theta\ns_r\, H^{(2r)}\Big]\quad, \label{eq:WWE-alpha}
\end{equation}
where $\alpha\in[0,1]$ and where the seven matrices $\{H^{(1)},\ldots,H^{(7)}\}$ are independently chosen from the GUE.
The corresponding structure factor is
\begin{equation}
\Hf\ns_{\alpha\ssr{-WWE}}(\Bvth)=1-\alpha + {\alpha\over 3} \big(\cos\vth\ns_1 + \cos\vth\ns_2 + \cos\vth\ns_3\big)\quad,
\end{equation}
which is again normalized.  At $\alpha=0$, there is no dependence on the parameters $(\theta\ns_1,\theta\ns_2,\theta\ns_3)$ and
no degeneracies are induced by varying $\Btheta$.  All the Chern numbers are zero.  As $\alpha$ increases toward $\alpha=1$ and 
we approach the WWE, monopoles appear, corresponding to degeneracies between neighboring bands, which lead to increases and decreases 
of Chern number.

From the WWE parametric random matrix model, we may compute the statistics 
of the Weyl points and Chern numbers, such as the densities $\rho_n^\pm(\Btheta)$ and $\lambda^\pm_n(\theta\ns_3)$
and the corresponding correlation functions $\blangle \rho_n^\sigma(\Btheta)\,\rho_{n'}^{\sigma'}(\Btheta')\brangle$ and $\blangle\lambda^\sigma_n(\theta\ns_3)\,\lambda^{\sigma'}_{n'}
(\theta'_3)\brangle$.  The Chern number fluctuations are given by eqn. \ref{DeltaCn}.
We consider the fluctuations of $\blangle [\Delta C_n(\chi)]^2\brangle$ and $\blangle [\Delta S_n(\chi)]^2\brangle$ with respect to $\chi$ (the integration limit of $\theta_3$).
Furthermore, we expand on the relation between the Weyl point distribution and Chern number fluctuation mentioned above, since they can serve as consistency check for our numerical data and analytical calculations. 

Computation of the Chern number fluctuations thus requires the knowledge of the monopole correlations, which will be presented in later sections. Here we discuss some limiting cases that can be readily computed. 
For small $\chi$ limit, it has been shown that the fluctuation behaves like random walk with linear dispersion,
\begin{equation}
\blangle[\Delta C_n(\chi)]^2\brangle =  {1\over \pi}\big(N^\pm_n+N^\pm_{n-1}\big)\,\chi+\CO(\chi^2)\quad,
\end{equation}
and 
\begin{equation}
\blangle\big[\Delta S\ns_n(\chi)\big]^2\brangle = {1\over\pi}\,N^\pm_n\,\chi+\CO(\chi^2)\quad,
\end{equation}
with $N_n^\sigma=\iiint\rho_n^\sigma(\Btheta)\>d^3\theta$ the total number of polarity $\sigma$ Weyl points between bands 
$n$ and $n+1$.  Note $N_n^+=N_n^-\equiv N^\pm_n$.  As discussed by Walker and Wilkinson \cite{walker1995universal},
the Weyl point correlations only make higher order contributions to the fluctuation for small $\chi$.
The derivation is reproduced in Appendix.\ref{sec:Chern_deg_+-_loc} for convenience.
For larger $\chi$, the correlations would make finite contributions and the fluctuation deviates from linearity. 
As a reference to contrast with numerical data, we compute the fluctuation with Weyl points treated as if independently distributed. 
The result is shown in Appendix.\ref{sec:Chern_deg_+-_loc} to be a parabola:
\begin{equation}
\blangle \left[\Delta C_n(\chi)\right]^2\brangle = {1\over\pi}\big(N^\pm_n + N^\pm_{n-1}\big)\,\chi\,
\Big(1-{\chi\over 2\pi}\Big)\qquad,\qquad 
\blangle \left[\Delta S_n(\chi)\right]^2\brangle = {1\over\pi}\,N^\pm_n\,\chi\,\Big(1-{\chi\over 2\pi}\Big)\quad.
\label{eq:delC_random_walk}
\end{equation}
This starts out linearly for small $\chi$ as a random walk, 
and eventually returns to $\langle(\Delta C\ns_n)^2\rangle=0$ at $\chi=2\pi$ because of periodicity. 
The presence of correlations will manifest in deviation from this result. 

\subsection{Method of locating Weyl points\label{subsec:find_deg}}

To locate the Weyl points, we divide the three dimensional parameter
space of $\left(\theta_{1},\theta_2,\theta_3\right)$ into fine enough cells
so that most of them contain at most a single Weyl point. The
Berry curvature flux of each cell is computed, and the presence and
polarity of the Weyl point is detected by the nonzero flux and
its sign, respectively. We adopt the algorithm of \cite{fukui2005chern}
for the numerical method, where the flux through each plaquette is
computed by the logarithm of the Wilson loop of around it, which guarantees
the computed flux through a closed surface to be an integer and converges
for modest grid resolution. The fluctuation of the Chern numbers is
then obtained from the results of \S \ref{WPCN}. 

\section{Numerical Result}\label{sec:numerical}

In this section, we present numerical results of the Chern numbers fluctuation 
and the correlation of Weyl points. 
Our data shows that the Chern number fluctuation saturates at a plateau after initial linear growth, 
and there is short-range correlation between opposite polarity Weyl points. 
We also find a scaling relation of the saturation level of the Chern number fluctuation with respect to the total number of bands, 
which complements the scaling of the number of Weyl points discussed in previous work \cite{walker1995universal}. 

\subsection{Diffusive behavior of the Chern number}

\begin{figure}[h]
\centering
\begin{subfigure}[t]{0.45\textwidth}
    \centering
    \includegraphics[width=\textwidth]{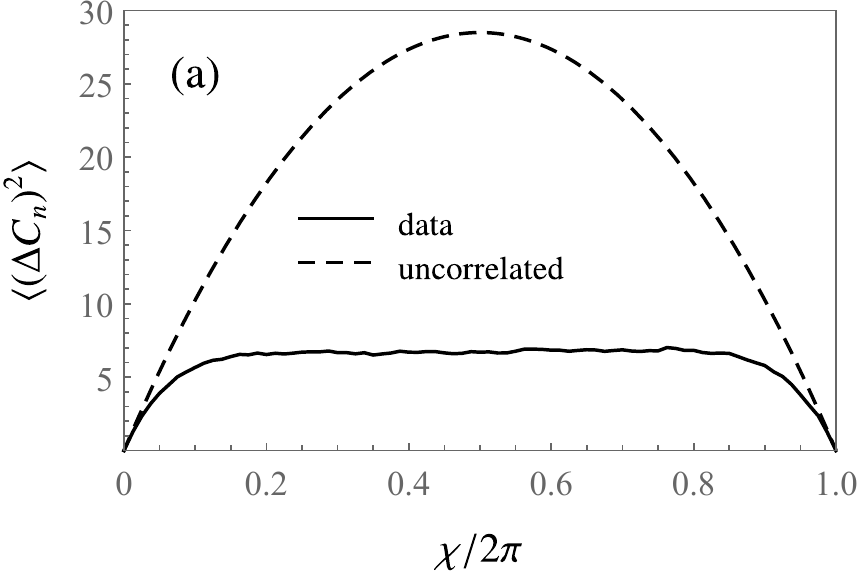}
\end{subfigure}
~~
\begin{subfigure}[t]{0.45\textwidth}
    \centering
    \includegraphics[width=\textwidth]{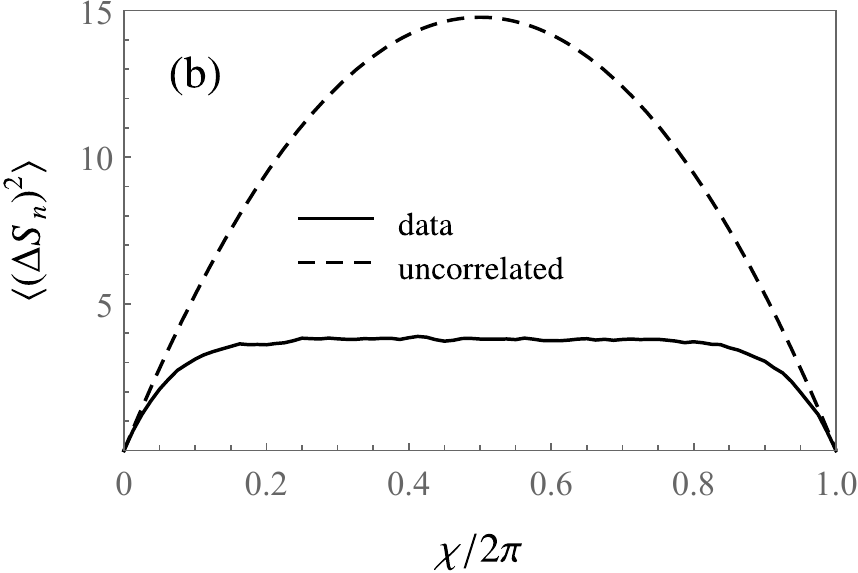}
\end{subfigure}
\caption{The fluctuations (a)  $\langle \Delta C_n(\chi)^2\rangle $ 
and (b) $\langle \Delta S_n \left( \chi \right) ^2 \rangle $ for the parametric GUE model
(solid line) compared with the hypothetical case where the correlations between the Weyl points are ignored (dashed). The data shown are obtained from the $n=4$ band, where the total number of bands $M=10$. \label{fig:dC_GUE}}
\end{figure}

As shown in Fig. \ref{fig:dC_GUE}, 
the fluctuation of Chern numbers $(\Delta C_n)^2$ and $(\Delta S_n)^2$
are evidently different from the parabola in (\ref{eq:delC_random_walk}) where correlations are ignored.
While the fluctuations do start linearly at small $\chi$, they eventually
saturate at a plateau\footnote{The eventual drop back to 0 at $\theta_3=2\pi$ is due to the periodicity
of the parameter space}. 
This indicates the presence of correlation between Weyl points, which is discussed below. 
In addition to the data shown here, we have checked other energy bands and different system sizes with total number of bands from $M=10$ to 60. 
They all exhibit the same behavior, only with different plateau height and the value of $\theta_3$ at which the fluctuation saturates. 

\subsection{Correlation of Weyl points}
\begin{figure}[!t]
\centering
\begin{minipage}[t]{0.45\textwidth}
    \centering
    \begin{subfigure}{\textwidth}
    \centering
    \includegraphics{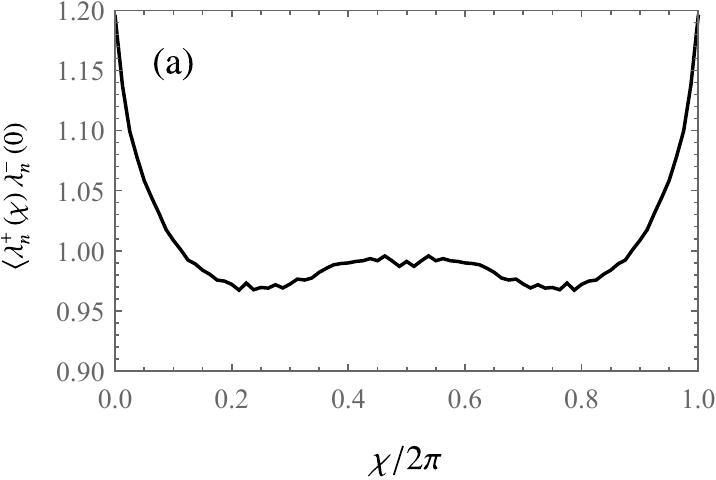}
    \phantomcaption{\label{fig:ffunc_M10_n4_+-}}
    \end{subfigure}
    \\ \par\bigskip
    \begin{subfigure}{\textwidth}
        \centering
        \includegraphics{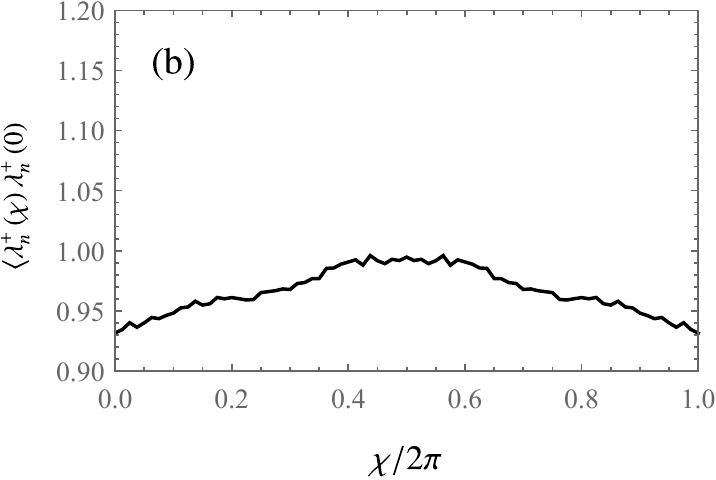}
        \phantomcaption{\label{fig:ffunc_M10_n4_++}}
    \end{subfigure}
    \\ \par\bigskip
    \begin{subfigure}{\textwidth}
        \centering
        \includegraphics{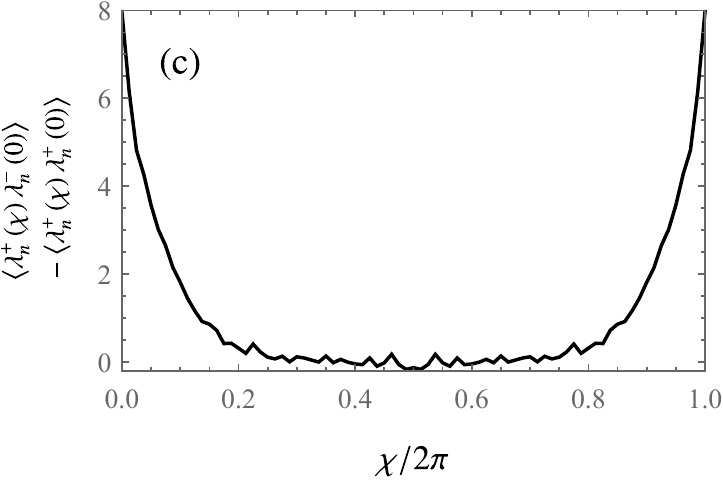}
        \phantomcaption{\label{fig:ffunc_M10_n4_+--++}}
    \end{subfigure}
    \caption{The correlation functions of 
    (a) opposite-polarity points $\blangle\lambda_n^+(\chi)\,\lambda_n^-(0)\brangle $, 
    (b) same-polarity points $\left\langle \lambda_n^+(\chi)\,\lambda_n^+(0)\right\rangle $, 
    and (c) the difference  $\left\langle \lambda_n^+(\chi)\,\lambda_n^+(0)\right\rangle -\blangle \lambda_n^+(\chi)\,\lambda_n^-(0)\brangle $, 
    normalized so that they integrate to one.
    Data drawn from band $n=4$ and total number of bands $M=10$. \label{fig:ffunc_M10_n4}}
\end{minipage}\hfill
\begin{minipage}[t]{0.45\textwidth}
    \centering
    \begin{subfigure}{0.95\textwidth}
        \centering
        \includegraphics{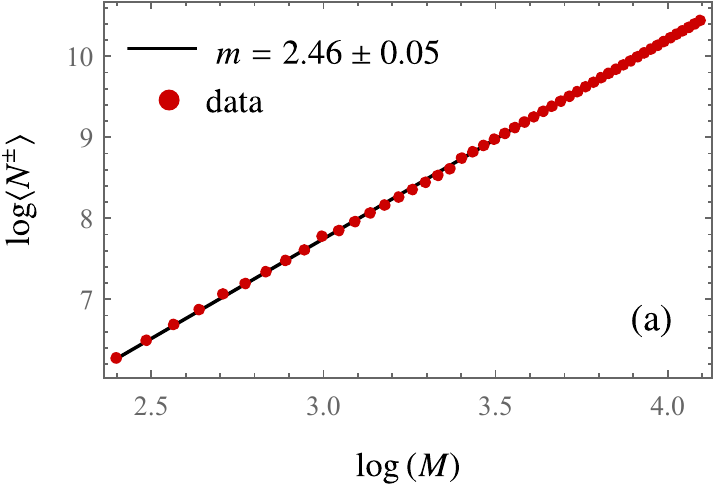}
        \phantomcaption{\label{fig:g_M_scaling}}
    \end{subfigure}
    \\ 
    \begin{subfigure}{0.95\textwidth}
        \centering
        \includegraphics{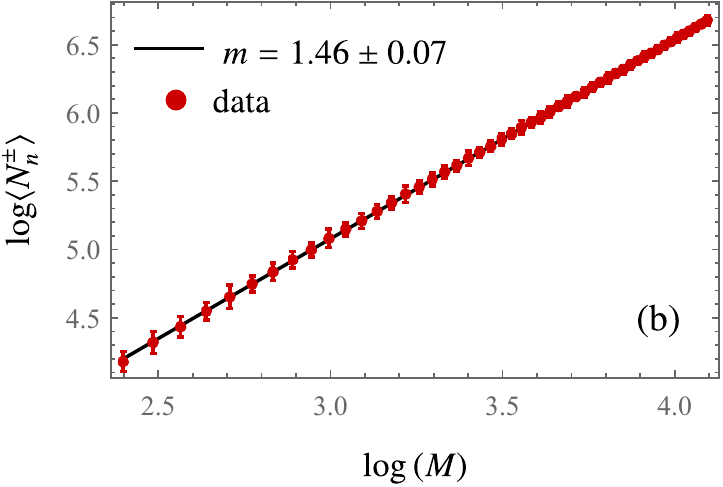}
        \phantomcaption{\label{fig:g_n0.4_scaling}}
    \end{subfigure}
    \\ 
    \begin{subfigure}{0.95\textwidth}
        \centering
        \includegraphics{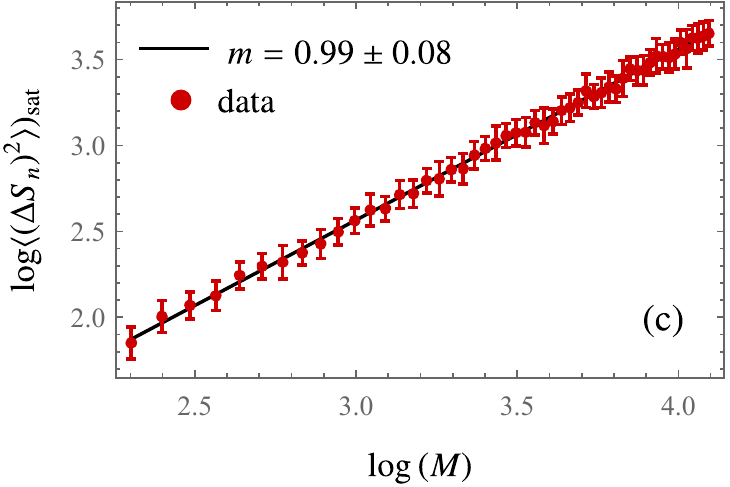}
        \phantomcaption{\label{fig:Cmax_scaling}}
    \end{subfigure}
    \caption{The scaling of (a) total number of Weyl points for all levels $\langle N^{\pm} \rangle \equiv \sum_n \langle N_n^{\pm} \rangle$, (b) the number of Weyl points for a single level $\langle N_n^{\pm} \rangle$, and (c) the plateau height for a single level $\blangle \big[S_n(\chi)\big]^2\brangle\nd_{\rm sat}$, versus the total number of bands
f    size $M$. 
    The error bars in (a) are smaller than the data dots and not shown. 
    The single levels are chosen at a fixed level with respect to the total number of bands, $n=0.4M$.  \label{fig:scaling_numerical}}
\end{minipage}
\end{figure}

We show the one-dimensional correlation function $\langle \lambda_n^\sigma(\theta\ns_3)\,\lambda^{\sigma'}_{n'}(\theta'_3)\rangle$
as it is most directly linked to the Chern number as discussed in \S \ref{WPCN},
where $\lambda_n^\sigma(\theta\ns_3)$ is the linear monopole density defined in (\ref{lde}). 
The correlation
between opposite-polarity and same-polarity Weyl points are shown in Fig.\ref{fig:ffunc_M10_n4}. 
There is evident short-range correlation in the former, which is preserved in the difference $\left\langle \lambda_n^+(\theta\ns_3)\,\lambda_n^+(0)\right\rangle -\left\langle\lambda_n^+(\theta\ns_3)\,\lambda_n^-(0)\right\rangle$, 
that directly leads to the Chern number fluctuation $(\Delta C\np_n)^2$, 
also shown in Fig. \ref{fig:ffunc_M10_n4}. 
On the other hand, while correlations for Weyl points between different bands $\blangle\rho_n^\sigma(\Btheta)\,\rho_{n'}^{\sigma'}(\Btheta')\brangle $
also contribute to the single level Chern number $C_n$, 
however our data in Appendix.\ref{sec:corr_across} shows that for the different band correlation, the same polarity and opposite polarity pieces cancel each other, making no contribution to the Chern number fluctuation. 

\subsection{Scaling relations}
Several scaling relations can be extracted from our numerical data.
First, we reproduce the previously known scaling relation
of the total number of Weyl points with respect to the total number of bands $M$
\cite{walker1995universal}, $\sum_{n=1}^{M-1} N_n^{\pm} \propto M^{5/2}$ shown in Fig. \ref{fig:g_M_scaling}. 
The average number of Weyl points per level should then scale
as $N_n^\pm\propto M^{3/2}$, and one might suspect that this
scaling also applies to the number for a fixed band $g_n$, 
which is indeed verified in Fig. \ref{fig:g_n0.4_scaling}. 
Moreover, we find another scaling relation for the height of the Chern number fluctuation plateau, 
$\blangle \big[S\nd_n(\chi)\big]^2\brangle\nd_{\rm sat}\propto M^1$,
as shown in Fig. \ref{fig:Cmax_scaling}. This also reflects the scaling relation for the correlation 
length, which we discuss in the following section. 
To make a fixed band index $n$ well-defined in the large $M$ limit, we choose the one at a fixed ratio to the total number of bands, i.e.~$r=n/M$ fixed, although the scaling behaviors we discussed are insensitive to the choice of the ratio $r$. 

\section{Analytical interpretation}\label{sec:analytical}

Having presented the numerical result, we discuss several different ways, in increasing rigor, to understand the short-range correlation between Weyl points and the fluctuation plateau. 
First, a qualitative picture which connects the two is illustrated; 
then the scaling relation of the correlation length is computed by perturbation theory, leading to the scaling exponent of the fluctuation plateau height. 
Finally, the entire functional form of the Chern number fluctuation is derived under the assumption that the short-range correlation function is exponential, which fits the data well. 

\subsection{Short-range correlation results in the saturation of the Chern number fluctuation\label{subsec:corr_intuitive}}

\begin{figure}
\centering
\begin{subfigure}{\textwidth}
    \centering
    \hspace{0.4cm}\includegraphics[scale=0.7]{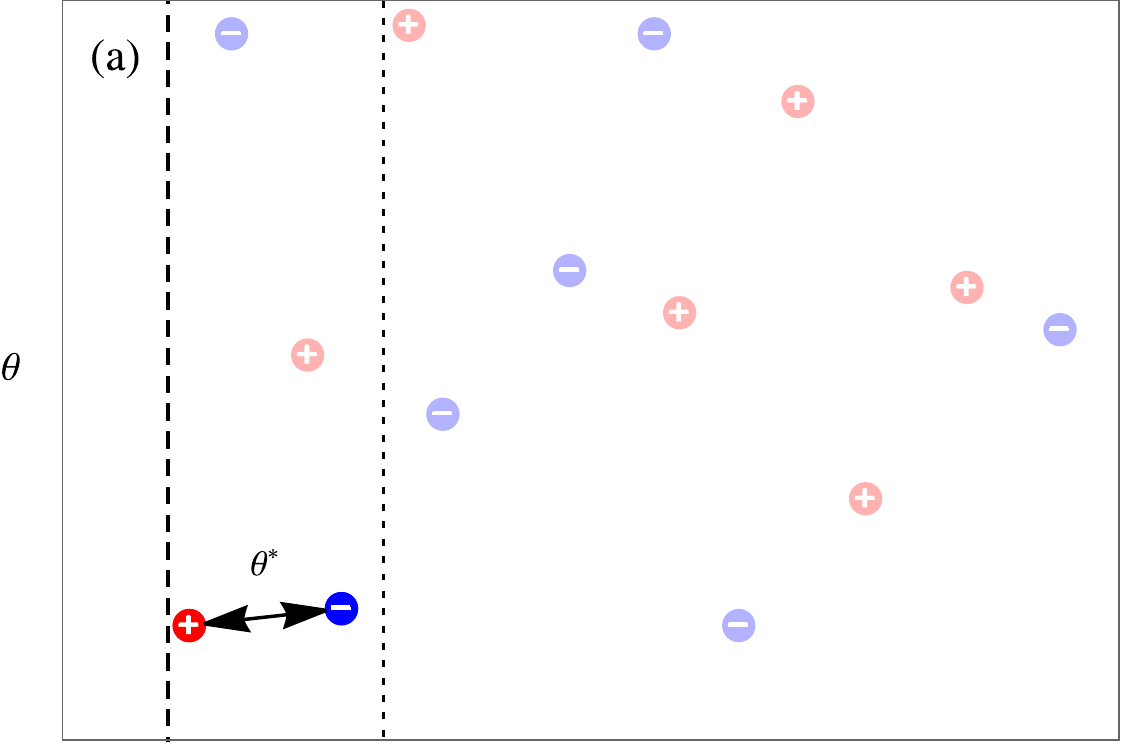}
    \phantomcaption{\label{fig:deg_2d_corred_labeled}}
\end{subfigure}
\\
\begin{subfigure}{\textwidth}
    \centering
    \includegraphics[scale=0.7]{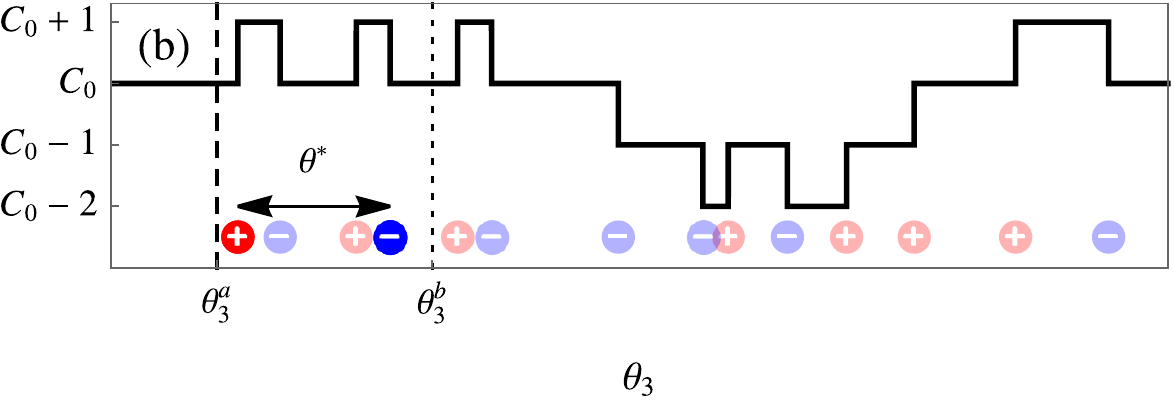}
    \phantomcaption{\label{fig:deg_1d_corred_labeled}}
\end{subfigure}
\caption{Correlated opposite polarity Weyl points shown as blue and hollow points at $\kappa$ apart, with other uncorrelated
ones shown in lighter colors. The bottom plot shows the one-dimensional distribution and
the change in the Chern number. \label{fig:k3_progress_corr}}
\end{figure}

Here we provide a qualitative picture showing that the short-range correlation between
opposite-polarity Weyl points results in the saturation plateau of Chern number fluctuation.
Because of the short-range correlation, an opposite-polarity point would likely be present within the correlation length $\kappa$ apart from a particular Weyl point, while farther Weyl points are uncorrelated, as illustrated in Fig. \ref{fig:k3_progress_corr}. 
Now consider a $\theta\ns_3$ plane and compare the Chern number $C(\theta\ns_3)$ to that on another plane 
$C(\theta'_3)$, where  the difference is contributed by the Weyl points located in the region between the two planes. 
Increasing the separation $\Delta \theta_3$ from zero adds new points to the region and changes the Chern number.
For $\Delta \theta\ns_3 \equiv \theta'_3 - \theta\np_3 \ll \kappa$, most of the the short-range correlated points still locate outside the region, and the included points are uncorrelated. 
The fluctuation thus follows a random walk, growing linearly with the number of Weyl points. 
If we project the Weyl points to focus only on the $\theta_3$ coordinate, as shown in Fig. \ref{fig:deg_1d_corred_labeled}, 
the points within $\Delta \theta_3 \ll \kappa$ are in fact separated in the $\theta_1$, $\theta_2$ direction and uncorrelated. 
When $\Delta \theta_3 \gg \kappa$, the correlated opposite-polarity points will be added to the region,
which cancels with existing contribution, so the fluctuation ceases to grow\footnote{For each newly added point, a short-range correlated point may be located either within or outside the region in between the plains of equal likelihood, the former case cancels out and reduces existing fluctuation while the latter contributes to it, so on average the fluctuation does not change.}. 
The saturated value is reached by the total linear growth from the $\Delta \theta_3 < \kappa$ range, 
determined by the number of Weyl points within this region:
\begin{equation}
\Big\langle \big(\Delta C_n\big)^2\Big\rangle _{\text{sat}}\propto  {1\over\pi}\,
\big(N^\pm_n+ N^\pm_{n-1}\big)\,\kappa\qquad,\qquad
\blangle(\Delta S\ns_n)^2\brangle\nd_{\rm sat}
\propto {1\over\pi}\,N^\pm_n\,\kappa\quad.
\label{eq:Cmax_scale}
\end{equation}

\subsection{Perturbation theory calculation of the scaling relations\label{subsec:perturb}}

We now compute the scaling coefficient of the correlation length using perturbation theory. 
Expanding around an existing Weyl point, we solve for other nearby Weyl points and compute the expected distance
$\kappa$ between them. 
This directly leads to the scaling coefficient of the fluctuation plateau height, which can be compared to our numerical data. 

Consider perturbing the Hamiltonian around a particular Weyl point, taken to be of positive polarity and located at $\Btheta=\left(0,0,0\right)$ without loss of generality. 
Assuming that the two bands are degenerate at zero energy (\ie\ the energy reference is always shifted to the degenerate level), the Hamiltonian at the Weyl point in the diagonal basis takes the following form, 
\begin{align}
H_0 & =\begin{pmatrix} 0_{2\times2} & 0_{2\times\left(M-2\right)} \\
0_{\left(M-2\right)\times2} & B_0 \end{pmatrix}\quad,\label{eq:H_pert_full}
\end{align}
where $B_0$ is a $\left(M-2\right)\times\left(M-2\right)$ diagonal matrix containing the the remaining energy bands. 
Away from the Weyl point, the Hamiltonian takes the general form of 
\begin{equation}
\begin{split}
H&= H_0 + \begin{pmatrix} F\nd_1 & A\yd_1 \\ A\nd_1 & B\nd_1\end{pmatrix} \theta\np_1 +
\begin{pmatrix} F\nd_2 & A\yd_2 \\ A\nd_2 & B\nd_2\end{pmatrix} \theta\np_2 +
\begin{pmatrix} F\nd_3 & A\yd_3 \\ A\nd_3 & B\nd_3\end{pmatrix} \theta\np_3 + \CO(\Btheta^2)\\
&=\begin{pmatrix} \BF\cdot\Btheta & \BA\yd\cdot\Btheta \\ \BA\cdot\Btheta & B\nd_0 + \BB\cdot\Btheta\end{pmatrix}
+\CO(\Btheta^2)\quad,
\end{split}
\end{equation}
where $F_{1,2,3}$ and $B_{1,2,3}$ are random Hermitian matrices
and $A_{1,2,3}$ are random complex rectangular matrices. 
By perturbation theory, we can obtain the effective Hamiltonian in the first $2\times2$ block, 
\begin{equation}
H_{\rm eff}= \BF\cdot\Btheta-(\Btheta \cdot \BA)\yd\,B_0^{-1}\, (\Btheta \cdot \BA)\quad,
\end{equation}
which may be expanded in the Pauli basis, {\it viz.\/}
\begin{equation}
H\ns_{\rm eff}= \big(v\ns_{ai}\,\theta\ns_i-c\ns_{aij} \,\theta\ns_i\, \theta\ns_j\big)\,\sigma^{a}\quad,
\label{eq:pert_Heff}
\end{equation}
with coefficients given by $v\ns_{ai}=\half\Tra\big(\sigma^a\,F\ns_i\big)$ and
$c\ns_{aij}=\half\Tra\big(\sigma^a\,A\yd_i\,B_0^{-1}\,A\nd_j\big)$.
We then solve $\det H(\Btheta) =0$ to obtain the roots $\Btheta=\Btheta^*$, which are the locations of the nearby
Weyl points.  These solutions $\Btheta^*$ must contain as many positive and negative Weyl points, 
since the Chern numbers must sum to zero for the two energy bands of $H_{\rm eff}$. 
One of them is the original degeneracy point $\Btheta^*=0$,
thus the rest must consist of exactly one more point with polarity opposite to that of the original one than ones with the  same polarity. 
From B{\'e}zout's theorem, there are then at most eight point solutions to the three second-order 
equations in the three variables $\Btheta^*$. However, numerical test shows that there are typically
only two or four such solutions.

As $v_{ai}$ does not scale with system size, the scaling of $\Btheta^*$  can be estimated from that of the 
coefficients $c_{aij}$, which is still difficult to compute. 
We therefore simplify $H_{\rm eff}$ by assuming that the only significant contribution to $B_0^{-1}$
in (\ref{eq:H_pert_full}) is from the closest nondegenerate band, with energy $E_{\rm nd}$, ignoring all other contributions. Then
\begin{equation}
H_{\rm eff}= \BF\cdot\Btheta - {(\BA\yd\cdot\Btheta)(\BA\cdot\Btheta)\over E\ns_{\rm nd}}\quad.
\end{equation}
Thus, $\kappa\propto 1/c\ns_{abc}\propto E\ns_{\rm nd}$\,.
Since the Hamiltonian is derived from the GUE ensemble, $E\ns_{\rm nd}$
follows Wigner's surmise \cite{mehta2004random}
\begin{equation}
P(s)=\frac{32}{\pi^2}\,s^2 e^{-4s^2/\pi}\quad,
\end{equation}
where $s=E\ns_{\rm nd}/D$, with $D$ mean distance between energy
levels. Assuming that $E\ns_{\rm nd}$ and the entries in $A$ are
independent, we have 
\begin{equation}
\textsf{var}(c^{-1})\propto\mathbb{E}\big[E^2_{\rm nd}\big]=
\half D^2\,\mathbb{E}(s^2)=\frac{3\pi}{16}\,D^2\quad,
\end{equation}
with $D=W/M$, $M$ the number of levels (the matrix size) and $W$
the spectral width. Under the semicircular law, $W=4\sqrt{M}$, so
$D=4/\sqrt{M}$ and $\textsf{var}(c^{-1})\propto M$,
leading to the scaling of the correlation length with respect to system size, $\kappa\propto  M^{-1/2}$.
Note that the only relevant scaling comes from the mean distance between
energy levels $D$, while the functional form of Wigner's surmise
simply yields a dimensionless constant.  Plugging this into (\ref{eq:Cmax_scale}),
we obtain $\blangle\big[\Delta S\ns_n(\chi)\big]^2\brangle\ns_{\rm sat}\propto M^1$
and similarly $\blangle\big[\Delta C\ns_n(\chi)\big]^2\brangle\ns_{\rm sat}\propto M^1$,
which agrees with the numerical scaling exponent shown in Fig. \ref{fig:Cmax_scaling}, where we find 
numerically a scaling exponent of $0.99 \pm 0.08$. 

\subsection{Analytical derivation of the fluctuation plateau}

\begin{figure}
\begin{centering}
\includegraphics{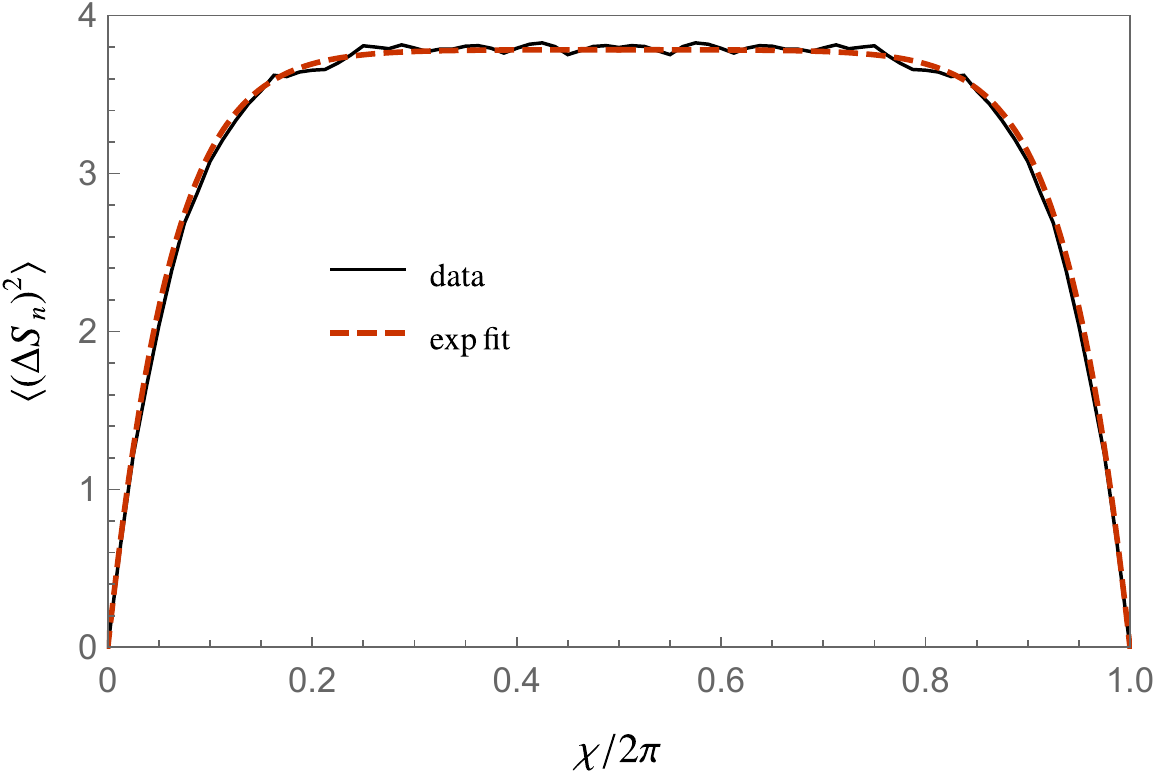}
\par\end{centering}
\caption{The Chern number fluctuation derived from
the exponential correlation function compared with data. \label{fig:delC_analytical_data} }
\end{figure}

Having computed the scaling relation for the correlation length, we calculate the Chern number fluctuation 
analytically for arbitrary $\Delta \theta_3$.  We assume that the short range correlation functions of 
$\blangle \rho_n^+ (\Bzero)\, \rho_n^-(\Btheta) \brangle$ decay exponentially, proportional to $e^{-|\Btheta|/\kappa}$
(or, more accurately, a periodic version thereof),
with the correlation length $\kappa=\kappa\nd_0\,M^{-1/2}$ following the scaling relation $\kappa\propto  M^{-1/2}$. 
Plugging this into our previous expressions, we obtain the closed form result for 
$\blangle\big[\Delta S\ns_n(\chi)\big]^2\brangle$, derived in Appendix \ref{sec:Chern_deg_+-_loc}:
\begin{align}
\blangle \big[\Delta S\np_n(\chi)\big]^2\brangle &\approx -
{\langle N^\pm_n\rangle\over\gamma\np_n}\,\Bigg\{\big[2\chib - 3 + (3+\chib)\,e^{-\chib}\big] +
{2\over \exp(\gamma\np_n) -1} \, \Big( 1 + {\gamma\np_n\over 1-\exp(-\gamma\np_n)}\Big) \big[\cosh\chib - 1\big]\label{Snchi}\\
&\hskip 2.5in - {2\over \exp(\gamma\np_n) -1}\>\big[\chib\sinh\chib - 2\cosh\chib + 2\big]\Bigg\}
+\langle N^\pm_n\rangle\,{\chi\over\pi} \quad,\nonumber
\end{align}
where $\chib=\chi/\kappa\np_n$ and $\gamma\np_n=2\pi/\kappa\np_n$.
The result agrees with the numerical data well, as shown in Fig. \ref{fig:delC_analytical_data}. 
The limits $\chi \ll \kappa$ and $\chi \gg \kappa$ can also be found, which shows linear growth of random walk and saturation plateau, respectively, 
\begin{equation}
\blangle\big[\Delta S\ns_n(\chi)\big]^2\brangle\approx\begin{cases}N^\pm_n\chi/\pi & {\rm if}\ \chi\ll \kappa\\
3N^\pm_n\kappa/2\pi & {\rm if}\ \chi\gg \kappa\quad.\end{cases}
\end{equation}
The saturation value is obtained in the $\kappa\ll 1$ limit from eqn. \ref{Snchi}.
The scaling of the plateau height can be obtained from that of the number of degeneracies, $N\ns_n=g\nd_0\,M^{3/2}$, 
and the correlation length $\kappa=\kappa\nd_0\,M^{-1/2}$, where $g\nd_0$ and $\kappa\nd_0$ are constants.  We then have
\begin{equation}
\blangle\big[\Delta S\ns_n(\chi)\big]^2\brangle\nd_{\rm sat}={3\over 2\pi}\,g\ns_0\,\kappa\ns_0\,M\quad,
\end{equation}
which agrees with the linear scaling with the total number of bands $M$ of our data in Fig. \ref{fig:Cmax_scaling}. 
For the single level Chern number $\left< (\Delta C_n)^2 \right>$, the fluctuation involves correlations across different bands $\blangle \lambda_n^\sigma(\theta\ns_3)\,\lambda_{n-1}^\sigma(\theta'_3)\brangle$.
Though we do not have an analytical understanding of this term, 
our numerical data in Appendix. \ref{sec:corr_across} shows that its net contribution to the fluctuation is negligible, 
so we arrive at a similar expression
\begin{equation}
\blangle\big[\Delta C\ns_n(\chi)\big]^2\brangle\approx \begin{cases}
\big(N^\pm_n+N^\pm_{n-1}\big)\chi/\pi & {\rm if}\ \chi\ll \kappa\\
3\big(N^\pm_n+N^\pm_{n-1}\big)\kappa/2\pi & {\rm if}\ \chi\gg \kappa\quad.\end{cases} \quad. 
\label{eq:delC_limit}
\end{equation}
That the analytical result fits the numerical data well supports the validity of our qualitative picture and the perturbation theory calculation. 
Other ansatzes for the correlation function are also considered in Appendix.\ref{sec:Chern_deg_+-_loc},
among which the exponential decay considered here produces the best fit. 

\section{Result for the one-parameter extension of Walker-Wilkinson ensemble}

\begin{figure}[hp]
    \centering
    \hspace{0cm}\includegraphics[scale=1]{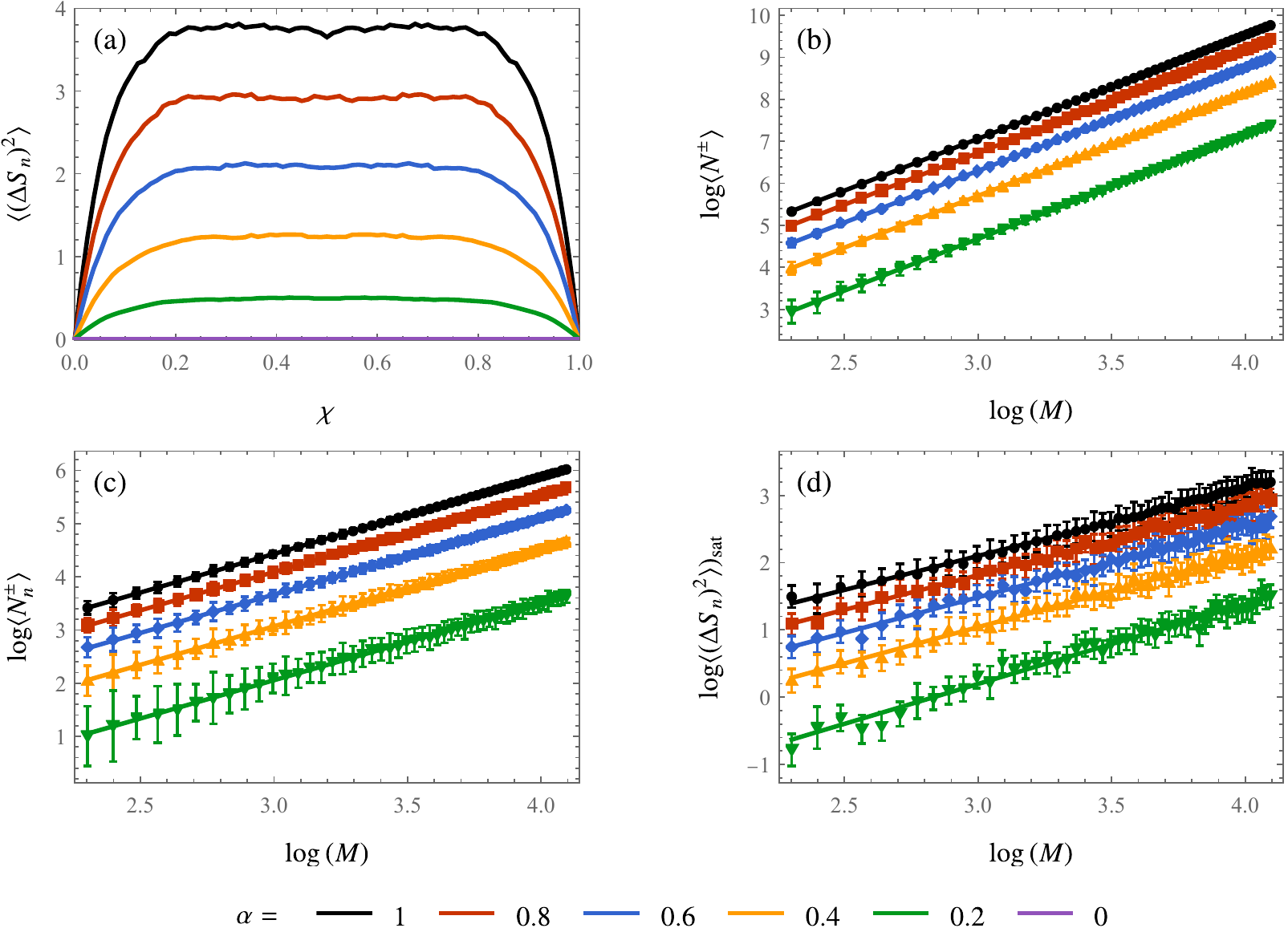}
    \caption{Numerical results for the one-parameter extension of WWE discussed in \ref{eq:WWE-alpha}, with (a) the fluctuation of Chern numbers for the level $n=4$ of system size $M=10$, and the scaling w.r.t. system size of (b) total number of degeneracy points, (c) number of degeneracy points at a single level, and (d) the saturation of Chern number fluctuation. The scaling exponents for different $\alpha$, shown in Table. \ref{tab:alpha}, agrees with $\alpha=1$. }
    \label{fig:alpha}
    \bigskip
    \begin{tabular}{|c|c|c|c|c|c|}
        \hline
        \backslashbox{quantity}{$\alpha$} & 1 & 0.8 & 0.6 & 0.4 & 0.2 \\
        \hline
         $\langle N^{\pm} \rangle$ & $2.47\pm0.04$ & $2.46 \pm 0.04$ & $2.46 \pm 0.05$ & $2.47 \pm 0.06$ & $2.47 \pm 0.09$ \\
        \hline
         $\langle N^{\pm}_n \rangle$ & $1.45 \pm 0.07$ & $1.44 \pm 0.08$ & $1.44 \pm 0.09$ & $1.45 \pm 0.1$ & $1.44 \pm 0.2$ \\
        \hline
         $\langle \left( \Delta S_n \right)^2 \rangle$ & $1.01 \pm 0.1$ & $1.03 \pm 0.1$ & $1.06 \pm 0.1$ & $1.08 \pm 0.1$ & $1.19 \pm 0.1$ \\
        \hline
    \end{tabular}
    \captionof{table}{The scaling exponents for the number of degeneracy points and Chern number fluctuation plateau height with respect to the system size $M$, plot presented in Fig. \ref{fig:alpha}}
    \label{tab:alpha}
\end{figure}

For completeness, we also show the result for the one-parameter extension of WWE discussed in \ref{eq:WWE-alpha}, 
where the parameter $\alpha$ interpolates between the WWE and a Hamiltonian independent of the parameters. 
Fig. \ref{fig:alpha}
shows the numerical result, consistent with our prediction where tuning $\alpha$ from 1 to 0 reduces the number of degeneracy points and Chern number fluctuation. 
Note that the scaling coefficients remain the same for different $\alpha$, which can be traced back to the derivation in Sec.\ref{sec:analytical}, 
where the scaling coefficient comes solely from the semicircular law of the random matrix ensemble, and the one-parameter Hamiltonian still satisfies the Gaussian unitary ensemble statistics. 

\section{Discussion}\label{sec:discussion}

In this work, we extended the analysis of a random matrix model with three parameters from Walker and Wilkinson \cite{walker1995universal} and investigated the 
Weyl points distribution in the parameter space and the fluctuation of Chern numbers. 
Our numerical data shows that there is short-ranged correlation between opposite-polarity degeneracy points,
and that the Chern number fluctuation, known to linear disperse for small $\Delta \theta_3$ \cite{walker1995universal}, eventually saturates at a plateau. 
A scaling relation for the height of such plateau with respect to the total number of bands is also found, complementing the already
known scaling relation for the total number of Weyl points.
To explain this result, we first provided a qualitative argument that connected the short-range correlation to the saturation of the Chern number fluctuation. 
Then, perturbation theory is utilized to quantitatively compute the scaling exponent. 
Finally, postulating that the short-range correlation function is exponentially decaying, the Chern number fluctuation is analytically derived, which agrees with the numerical data. 
We also investigated a one-parameter family of models which interpolates between the Walker-Wilkinson ensemble and a constant Hamiltonian. 
Our numerics shows that the number of degeneracy points and Chern number fluctuation decreases when tuned away from the WWE, as expected. 

Our result can potentially be applied to Weyl semimetal materials with many Weyl points in the bulk. 
The opposite-polarity Weyl points would tend to pair up due to the short-range correlation, and the expected numbers of Fermi arc states would be similar throughout most of the surface Brillouin zone, indicated by the saturation of the Chern number fluctuation. 
The scaling relations also indicates how much one can increase the number of Weyl poins and Fermi arc states by enlarging the system size. 

Although the random matrix model adopted here is derived from the GUE ensemble, representing a generic system without any symmetry, 
our qualitative arguments and computation steps can be readily adapted to different random matrix ensembles. 
The Chern number fluctuation is related to the Weyl point correlations by simple integrals, 
and the perturbation theory calculation of the scaling can be modified to include the Wigner surmise for the appropriate ensemble.
An example of this has been done in \cite{barakov2021abundance}, where random matrices obeying the Bogoliubov-de-Gunnes mirror symmetry were considered to describe  Weyl points in multichannel Josephson junctions. 
The band-touching points of different ensembles may have a different co-dimension and are associated with different topological invariant such as the second Chern number.
We leave this direction to future work. 

\bibliographystyle{unsrt}
\bibliography{reference_Chern}

\appendix

\section{Chern numbers from the location of degeneracies \label{sec:Chern_deg_+-_loc}}

Here we expand on the properties of the Weyl point correlation function and its consequences for
Chern number fluctuations. The density $\rho_n^\sigma(\Btheta)$ is given by
\begin{equation}
\rho_n^\sigma(\Btheta)=\sum_{j=1}^{N^\sigma_n}\dtil(\Btheta-\Btheta^\sigma_{n,j})\quad,
\end{equation}
where $\Btheta^\sigma_{nj}$ is the location of the $j^{\rm th}$ Weyl point of polarity $\sigma$ between bands 
$n$ and $n+1$, and where $\dtil(\Btheta)=\sum_\Bl\delta(\Btheta-2\pi\Bl)$ is periodic under displacement of the
dimensionless wavevector $\Btheta$ by any $\Bl=(l\ns_1,l\ns_2,l\ns_3)$ is a triple of integers.
For a given instantiation of $H(\Btheta)$, the number of such Weyl points
we define to be $N_n^\sigma$; clearly $N_n^+=N_n^-\equiv N^\pm_n$.  However these numbers
may fluctuate within our matrix ensemble.  Thus this is a `grand canonical' formulation.  We then have
\begin{equation}
\begin{split}
\Delta C\nd_n(\chi)&=\iTT\!d^2 \theta\ns_\perp\!\int\limits_0^\chi\! d\theta\ns_3\>
\sum_\sigma \sigma\, \Big\{\rho_n^\sigma(\Btheta)-\rho_{n-1}^\sigma(\Btheta)\Big\}\\
\Delta S\nd_n(\chi)&=\sum_{j=1}^n C\nd_j(\chi)=\iTT\!d^2 \theta\ns_\perp\!\int\limits_0^\chi\!d\theta\ns_3
\ \sum_\sigma \sigma \,\rho_n^\sigma(\Btheta)\quad,
\end{split}
\end{equation}
where $\TT$ denotes the 2-torus $[0,2\pi]\times[0,2\pi]$.
$\Btheta$-space translational invariance of the one- and two-point distributions entails
\begin{equation}
\blangle\rho_n^\sigma(\Btheta)\brangle={\langle N^\sigma_n\rangle\over 8\pi^3}\qquad,\qquad
\blangle\rho_n^\sigma(\Btheta)\,\rho_{n'}^{\sigma'}(\Btheta')\brangle=F_{n,n'}^{\sigma,\sigma'}(\Btheta-\Btheta')\quad,
\end{equation}
and we may write
\begin{equation}
F_{n,n'}^{\sigma,\sigma'}(\Btheta-\Btheta')={1\over 8\pi^3}\>\bigg\{\langle N^\pm_n\rangle\,
\delta\ns_{\sigma,\sigma'}\,\delta\ns_{n,n'}\,\dtil(\Btheta-\Btheta') + R_{n,n'}^{\sigma,\sigma'}(\Btheta-\Btheta')\bigg\}\quad,
\end{equation}
where $R_{n,n'}^{\sigma,\sigma'}(\Bvth)$ is regular, with no $\delta$-function singularities
in $\Bvth=\Btheta-\Btheta'$.  Note the sum rule,
\begin{equation}
\iTTT\! d^3\vth\>R^{\sigma,\sigma'}_{n,n'}(\Bvth)=\blangle N^\sigma_n\,N^{\sigma'}_{n'}\brangle -
\blangle N^\sigma_n\brangle\>\delta\ns_{n,n'}\,\delta\ns_{\sigma,\sigma'}\quad,
\end{equation}
where $\TTT$ denotes the three-dimensional torus. Note that $R^{\sigma,\sigma'}_{n,n'}(\Bvth)$ is reflection symmetric
under $q\ns_\mu\to -q\ns_\mu$ for $\mu\in\{1,2,3\}$.  We thus have
\begin{equation}
\begin{split}
\blangle \big[\Delta C\np_n(\chi)\big]^2\brangle &=\Big(\langle N^\pm_n\rangle + \langle N^\pm_{n-1}\rangle\Big)\,{\chi\over\pi}+2\int\limits_0^\chi \! d\alpha\!\!\int\limits_0^{\chi-\alpha}
\!\!\! d\theta\np_3\> \Big(\Lap_{n,n}(\theta\ns_3)+\Lap_{n-1,n-1}(\theta\ns_3)-2\Lap_{n,n-1}(\theta\ns_3)\Big)\\
\blangle \big[\Delta S\np_n(\chi)\big]^2\brangle &= \langle N^\pm_n\rangle\,{\chi\over\pi}
+2\int\limits_0^\chi \! d\alpha\!\!\int\limits_0^{\chi-\alpha}\!\!\! d\theta\ns_3\>\Lap_{n,n}(\theta\ns_3)\quad,
\end{split}
\label{eq:delC_from_Lambda}
\end{equation}
where
\begin{equation}
\Lambda\np_{n,n'}(\Bvth)={1\over 2\pi}\sum_{\sigma,\sigma'}\sigma\sigma' \,R^{\sigma,\sigma'}_{n,n'}(\Bvth)
\qquad,\qquad \Lap_{n,n'}(\vth\ns_3)=\iTT\! d^2 \vth\ns_\perp \> \Lambda\np_{n,n'}(\Bvth\ns_\perp,\vth\ns_3) \quad.
\end{equation}
If we assume that $R^{\sigma,\sigma'}_{n,n'}(\Bvth)$ is independent of $\Bvth$, \ie\ the locations of
different Weyl points are completely uncorrelated, then from the sum rule
\begin{equation}
\iTTT \! d^3\vth\ \Lambda\np_{n,n'}(\Bvth)=-{1\over\pi}\,\langle N^\pm_n\rangle\,\delta\ns_{n,n'}
\end{equation}
we have the crude approximation
\begin{equation}
\Lambda^\textsf{nc}_{n,n'}(\Bvth)=-{1\over 8\pi^4}\,\blangle N^\pm_n\brangle\,\delta\ns_{n,n'}\quad,
\end{equation}
from which we recover (\ref{eq:delC_random_walk}).  Here the superscript $\textsf{nc}$ stands for `no correlations'.

\begin{figure}
\begin{centering}
\includegraphics[width=5in]{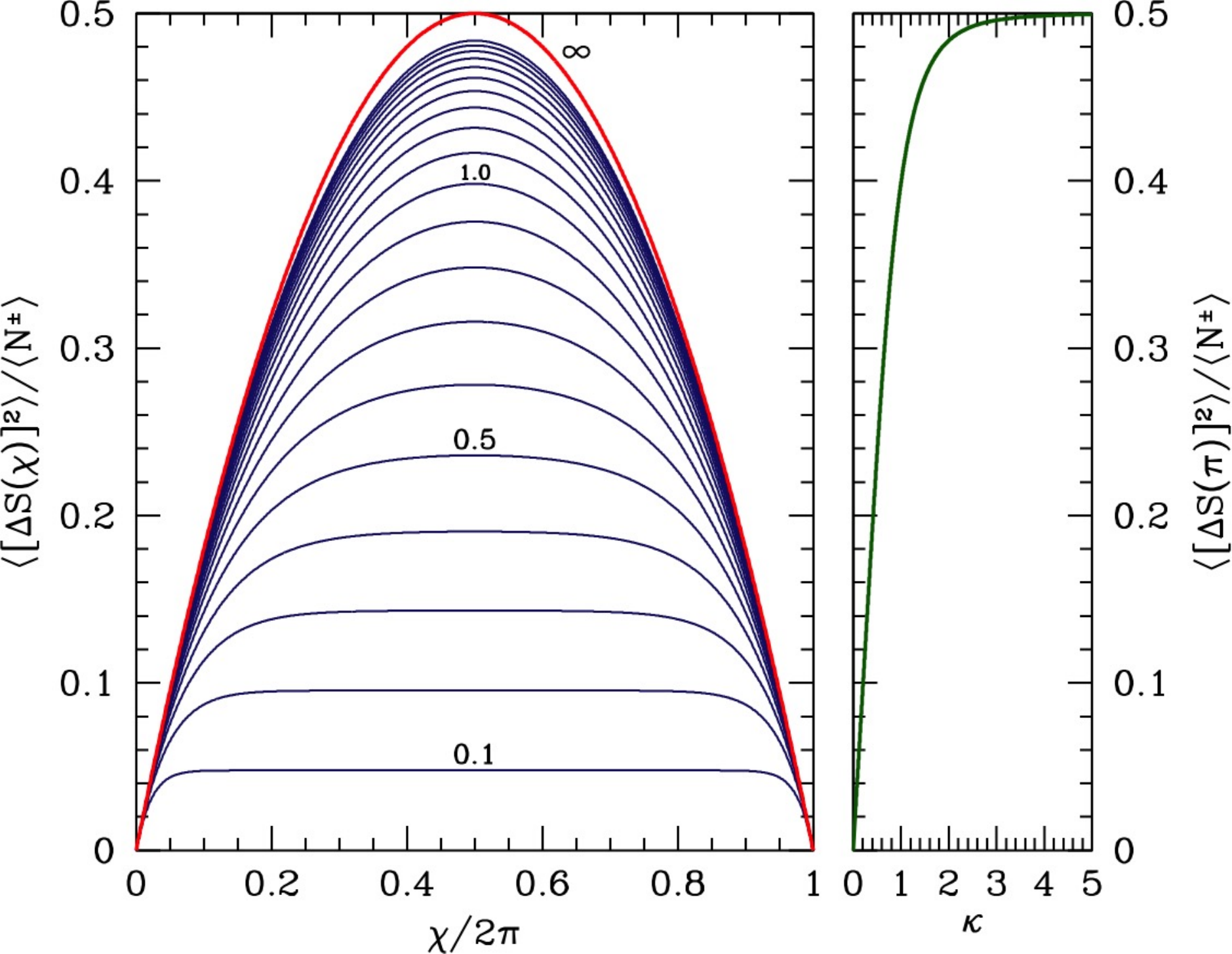}
\par\end{centering}
\caption{Predictions of the approximation formula eqn. \ref{Snchi} for the total Chern number 
fluctuation {\it versus\/} slab thickness $\chi$.  Left: results for 20 equally spaced values of $\kappa$
and $\kappa=\infty$.  Right: the values at $\chi=\pi$.}
\label{Sappfig}
\end{figure}

We now extend this analysis to include correlations among different Weyl points using a phenomenological
description.  Again imposing the sum rule constraint, we write
\begin{equation}
\Lambda\np_{n,n'}(\Bvth)\approx -\delta\np_{n,n'}\ {\blangle N^\pm_n\brangle\over 8\pi^2 \kappa_n^3}
\ \sum_{\Bl\in\MZ^3} e^{-|\Bvth-2\pi\Bl|/\kappa\np_n} \quad,
\end{equation}
where the sum on $\Bl$ is over all triples of integers.  Integrating over $(\vth\np_1,\vth\np_2)$, we find
\begin{equation}
\Lap_{n,n'}(\vth\ns_3) = -\,\delta\ns_{n,n'}\>{\blangle N^\pm_n\brangle\over 4\pi\kappa\np_n}
\sum_{l=-\infty}^\infty\bigg(1+{|\vth\ns_3-2\pi l|\over\kappa\np_n}\bigg)\,e^{-|\vth\ns_3-2\pi l|/\kappa\np_n}\quad.
\end{equation}
Our approximation results in
\begin{equation}
\blangle \big[\Delta S\np_n(\chi)\big]^2\brangle \approx \langle N^\pm_n\rangle\,{\chi\over\pi}
+2\int\limits_0^\chi \! d\alpha\!\!\int\limits_0^{\chi-\alpha}\!\!\! d\vth\ns_3\ \Lap_{n,n}(\vth\ns_3)
\end{equation}
and
\begin{equation}
\blangle \big[\Delta C\np_n(\chi)\big]^2\brangle \approx \blangle \big[\Delta S\np_n(\chi)\big]^2\brangle +
\blangle \big[\Delta S\np_{n-1}(\chi)\big]^2\brangle\quad.
\end{equation}

The sums and integrals can be performed, yielding the following closed-form expression:
\begin{align}
\blangle \big[\Delta S\np_n(\chi)\big]^2\brangle &\approx -
{\langle N^\pm_n\rangle\over\gamma\np_n}\,\Bigg\{\big[2\chib - 3 + (3+\chib)\,e^{-\chib}\big] +
{2\over \exp(\gamma\np_n) -1} \, \Big( 1 + {\gamma\np_n\over 1-\exp(-\gamma\np_n)}\Big) \big[\cosh\chib - 1\big]\label{Snchi}\\
&\hskip 2.5in - {2\over \exp(\gamma\np_n) -1}\>\big[\chib\sinh\chib - 2\cosh\chib + 2\big]\Bigg\}
+\langle N^\pm_n\rangle\,{\chi\over\pi} \quad,\nonumber
\end{align}
where $\chib=\chi/\kappa\np_n$ and $\gamma\np_n=2\pi/\kappa\np_n$\,.

\begin{figure}
\centering
    \centering
    \hspace{-0.7cm}
    \includegraphics{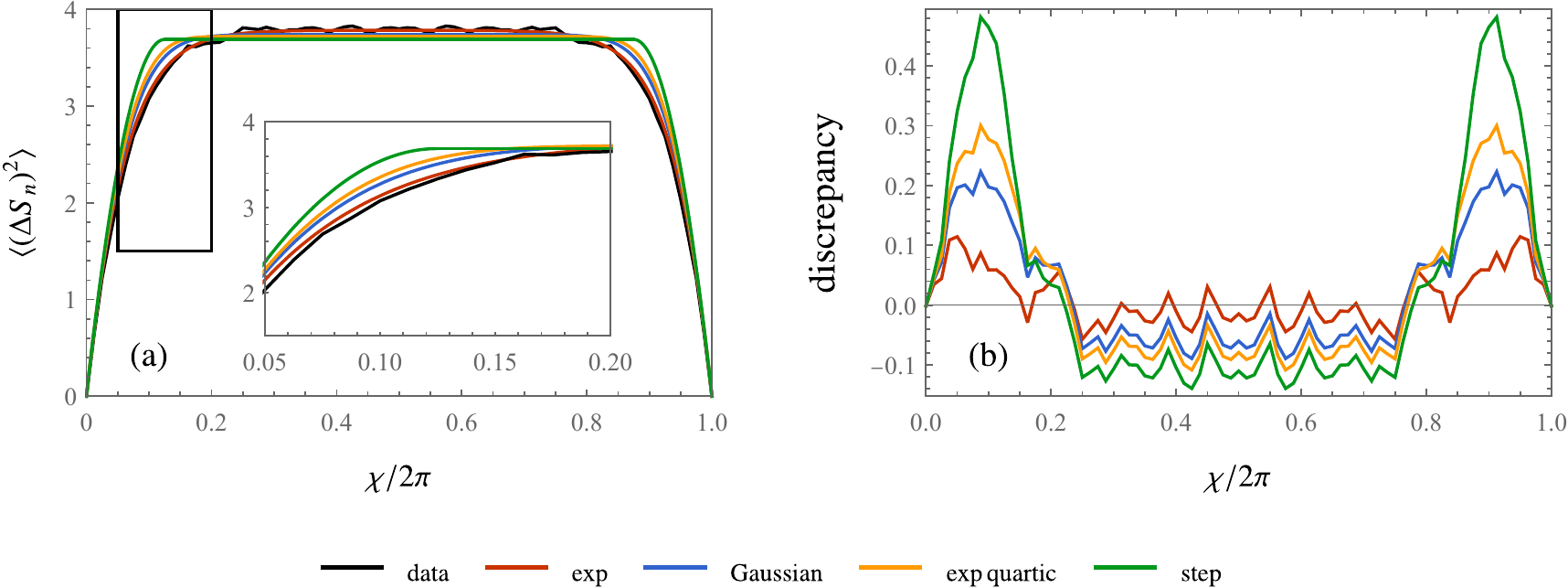}
\caption{The predicted Chern number fluctuation obtained from correlation functions
of different functional forms, (a) compared with data, and (b) the difference with data. 
The matrix rank is $M=10$.
\label{fig:deltaC_fit_all4-1} }
\end{figure}

For completeness we also tested different functional form for the
correlation function, including the Gaussian $\exp(-\Btheta^2/\kappa)$,
the quartic exponential $\exp\big(-|\Btheta|^4/\kappa\big)$,
and a hypothetical case where the one-dimensional correlation
function is a step function, obtaining the resulting Chern number
fluctuation. The results are shown in shown in Fig. \ref{fig:deltaC_fit_all4-1},
where the existence of the saturation plateau does not depend on the
correlation functional form, in agreement with the prediction in Sec.\ref{subsec:corr_intuitive},
and that the exponential function adopted in the main text fits the
numerical data best. 

\section{Special cases for the Chern number fluctuations}
\begin{figure}
\centering
\begin{subfigure}{0.45\textwidth}
    \centering
    \includegraphics[width=\textwidth]{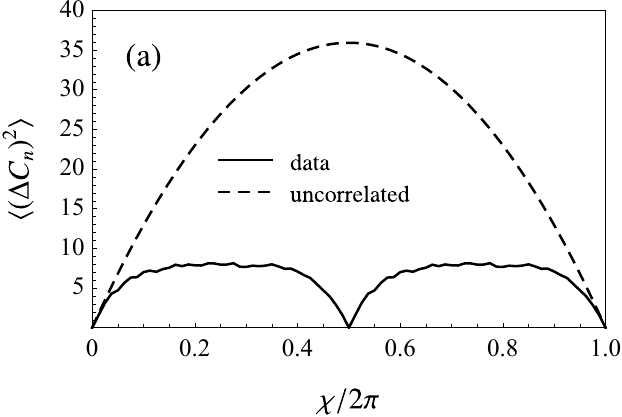}
    \phantomcaption
    \label{fig:delC_M11_n6}
\end{subfigure}
\hskip 0.3in
\begin{subfigure}{0.45\textwidth}
    \centering
    \includegraphics[width=\textwidth]{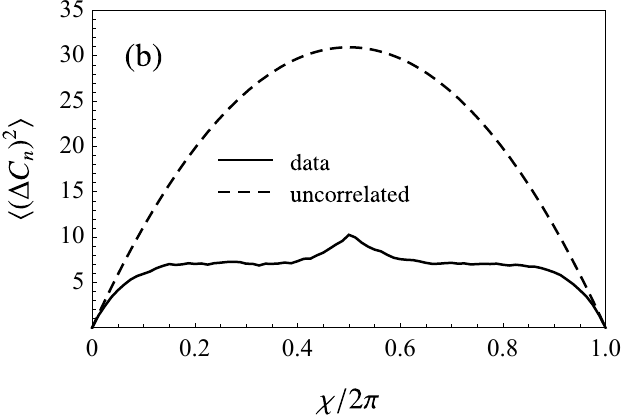}
    \phantomcaption
    \label{fig:delC_M10_n5}
\end{subfigure}
\caption{The Chern number fluctuation of the middle energy band for (a) odd total number of bands (b) even total number of bands. Data from  $n=6$ of
odd total number of bands $M=11$ and $n=5$ of
even total number of bands $M=10$, respectively. \label{fig:delC_middle}}
\end{figure}

Although the Chern number fluctuation for most energy band $n$ and different total numbers of bands 
$M$ obey the saturation pattern of Fig.\ref{fig:dC_GUE}, there are some special symmetries to our
$\Btheta$-parameterized GUE which constrain the correlation functions.  In addition to the periodicity
$H(\Btheta+2\pi\Bl)=H(\Btheta)$, where $\Bl=(l\ns_1,l\ns_2,l\ns_3)$ with each $l\ns_\mu\in\MZ$, we also have
$H(\Btheta+\Bpi)=-H(\Btheta)$, where $\Bpi=(\pi,\pi,\pi)$.  This entails the relation
$E\ns_n(\Btheta+\Bpi)=E\ns_{M+1-n}(\Btheta)$, and thus $\rho^\pm_n(\Btheta+\Bpi)=\rho^\mp_{M-n}(\Btheta)$.
In most cases this symmetry does not manifest in the Chern number fluctuations, since it does not
constrain the Weyl points for a given band.  The exceptions are the cases $n=M/2$ for $M$ even and
$n=(M+1)/2$ for $M$ odd, for which
\begin{equation}
\begin{split}
\rho^\pm_{M/2}(\Btheta+\Bpi)&=\rho^\mp_{M/2}(\Btheta) \qquad\quad\ \  \hbox{\rm ($M$ even})\\
\rho^\pm_{(M+1)/2}(\Btheta+\Bpi)&=\rho^\mp_{(M-1)/2}(\Btheta) \qquad \hbox{\rm ($M$ odd)}
\end{split}
\end{equation}
Integrating over $\theta\ns_{1,2}$ we have that
\begin{equation}
\begin{split}
\lambda^\pm_{M/2}(\theta\ns_3+\pi)&=\lambda^\mp_{M/2}(\theta\ns_3) \qquad\quad\ \  \hbox{\rm ($M$ even})\\
\lambda^\pm_{(M+1)/2}(\theta\ns_3+\pi)&=\lambda^\mp_{(M-1)/2}(\theta\ns_3) \qquad \hbox{\rm ($M$ odd)}
\end{split}
\end{equation}
for the one-dimensional distributions $\lambda_\nbar^\sigma(k)$, where $\nbar=\half M$ for $M$ even
and $\nbar=\half(M+1)$ for $M$ odd.  The Chern number change $\Delta C\ns_\nbar(\chi)$ is given in
eqn. \ref{DeltaCn}, and using the above results we have for $M$ odd that
\begin{equation}
\Delta C^{\rm odd}_\nbar(\chi)=\int\limits_0^\chi\! d\theta\ns_3\>\sum_\sigma\, \sigma\big(\lambda^\sigma_\nbar(\theta\ns_3) +
\lambda^\sigma_\nbar(\theta\ns_3+\pi)\big)\quad,
\end{equation}
and therefore
\begin{equation}
\Delta C^{\rm odd}_\nbar(\pi)=\int\limits_0^{2 \pi} \!d\theta\ns_3\>\sum_\sigma\sigma\,\lambda_\nbar^\sigma(\theta\ns_3)=0\quad.
\end{equation}
For $M$ even, we invoke 
\begin{equation}
\blangle \big[\Delta C\ns_n(\chi)\big]^2\brangle = \int\limits_0^\chi\!d\theta\ns_3\int\limits_0^\chi\!d\theta'_3\>
\sum_{\sigma,\sigma'}\sigma\sigma'\,\Big[\blangle\lambda_n^\sigma(\theta\np_3)\,\lambda_n^{\sigma'}(\theta'_3)\brangle
+\blangle\lambda_{n-1}^\sigma(\theta\np_3)\,\lambda_{n-1}^{\sigma'}(\theta'_3)\brangle - 2\,
\blangle\lambda_n^\sigma(\theta\np_3)\,\lambda_{n-1}^{\sigma'}(\theta'_3)\brangle\Big]
\end{equation}
and 
\begin{equation}
F^\parallel_{n,n'}(\theta\ns_3-\theta'_3)\equiv \sum_{\sigma,\sigma'}\sigma\sigma'\,\blangle\lambda^\sigma_n(\theta\np_3)\,\lambda^{\sigma'}_{n'}(\theta'_3)\brangle
={1\over \pi}\,\blangle N^\pm_n\brangle\>\delta\ns_{n,n'} \,\dtil(\theta\np_3-\theta'_3) 
+ \Lambda^\parallel_{n,n'}(\theta\np_3-\theta'_3)\quad,
\end{equation}
where $\dtil(\vth\ns_3)=\sum_l\delta(\vth\ns_3-2\pi l)$.  Invoking the symmetry under advancing 
$\vth\np_3=\theta'_3-\theta\np_3$ by $\pi$, we may write
\begin{equation}
F^\parallel_{\nbar,\nbar}(\vth\ns_3)={1\over \pi}\,\blangle N^\pm_\nbar\brangle\>\Big(\dtil(\vth\ns_3)-\dtil(\vth\ns_3-\pi)\Big)
+{\tilde\Lambda}^\parallel_{\nbar,\nbar}(\vth\ns_3) - {\tilde\Lambda}^\parallel_{\nbar,\nbar}(\vth\ns_3-\pi)\quad,
\end{equation}
where ${\tilde\Lambda}^\parallel_{\nbar,\nbar}(\vth\ns_3)$ is regular.  Integrating over $\theta\np_3$ and $\theta'_3$ to obtain
$\blangle \big[\Delta C\ns_n(\chi)\big]^2\brangle$, we have
\begin{equation}
\blangle \big[\Delta C\ns_\nbar(\chi)\big]^2\brangle = \blangle N^\pm_\nbar\brangle\bigg( 1 -
{|\chi-\pi|\over\pi}\bigg)+ \ldots\quad,
\end{equation}
where the first term on the RHS is due to the delta functions, and the terms included in the ellipses 
are smooth, periodic functions of $\chi$ which are symmetric about $\chi=\pi$.  Thus, there is a downward
cusp at $\chi=\pi$, as we observe in our simulations.  The results for $M$ even and $M$ odd are shown 
in Fig. \ref{fig:delC_M11_n6}.

We stress that the cusps present at $\chi=\pi$ for the special bands $\nbar$ in the even 
and odd $M$ cases are nongeneric features which arise due to the Walker-Wilkinson 
parameterization of $H(\Btheta)$ in eqn. \ref{eq:GUE_paramed}.

\section{Weyl point correlations among different energy bands\label{sec:corr_across}}

\begin{figure}
\begin{subfigure}{0.45\textwidth}
    \centering
    \includegraphics{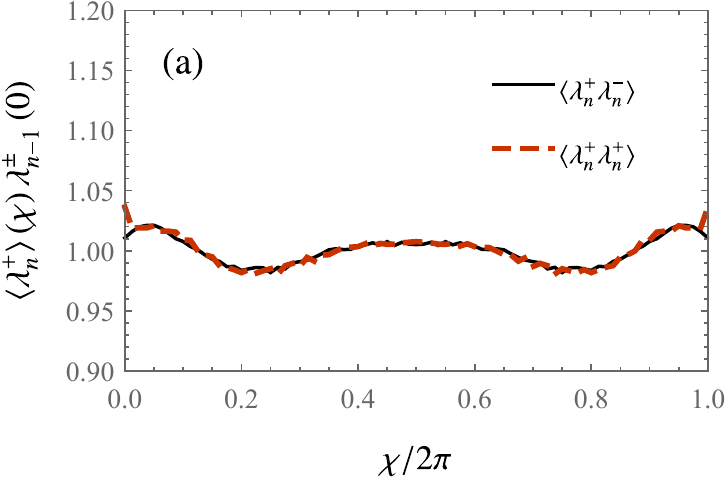}
    \phantomcaption{\label{fig:corr_across_++}}
\end{subfigure}
\hspace{0.2in}
\begin{subfigure}{0.45\textwidth}
    \centering
    \includegraphics{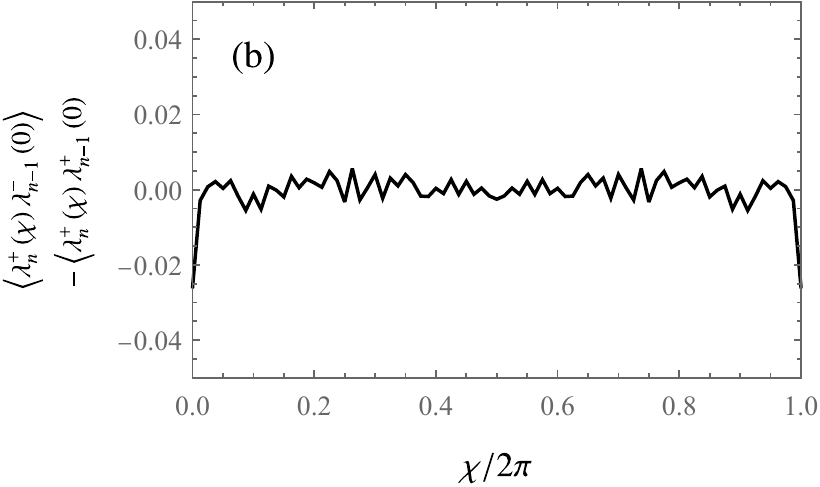}
    \phantomcaption{\label{fig:corr_across_+-}}
\end{subfigure}
\caption{The correlation among Weyl points between different energy bands. 
(a) same-polarity correlation $\blangle \lambda_n^+(\chi)\,\lambda_{n-1}^+(0)\brangle $ 
and opposite-polarity correlation $\blangle \lambda_n^+(\chi)\,\lambda_{n-1}^-(0)\brangle $, 
(b) the difference $\blangle \lambda_n^+(\chi)\,\lambda_{n-1}^+(0)\brangle -\blangle \lambda_n^+(\chi)\,\lambda_{n-1}^-(0)\brangle $. 
The first two are normalized to integrate to one, while the last is the difference of the former two without further normalization. 
Data is obtained for system size $M=10$ and $n=4$. \label{fig:corr_across}}
\end{figure}

In Fig. \ref{fig:corr_across} we show the numerical result for the correlations
of monopoles between different energy bands, 
$\blangle \lambda_n^\sigma(\chi)\,\lambda_{n-1}^{\sigma'}(0)\brangle $. 
which contribute to the fluctuation $\langle \left( \Delta C_n \right)^2 \rangle$ in \eqref{eq:delC_from_Lambda}
but cannot be quantified by our perturbation description. 
Here, the numerical data shows that the difference $\blangle \lambda_n^+(\chi)\,\lambda_{n-1}^+(0)\brangle -\blangle \lambda_n^+(\chi)\,\lambda_{n-1}^-(0)\brangle$ effectively cancel out, 
making a negligible contribution to $\langle \left( \Delta C_n \right)^2 \rangle$. 
Therefore, the behavior of $\langle \left( \Delta C_n \right)^2 \rangle$ can be to good accuracy calculated from our perturbation theory analysis, 
as in \eqref{eq:delC_limit}, despite our neglect of correlation effects across different energy bands. 

\end{document}